\documentclass[journal,onecolumn]{IEEEtran}
\usepackage[english]{babel}

\usepackage{cite}
\usepackage{epstopdf}

\ifCLASSINFOpdf
\else
\fi
\usepackage[T1]{fontenc}
\usepackage[latin9]{inputenc}
\usepackage{amsmath}
\usepackage{amsthm}
\usepackage{amssymb}
\usepackage{stmaryrd}
\usepackage{graphicx}
\usepackage{color,xcolor}
\usepackage{epstopdf}
\newtheorem{thm}{Theorem}
\newtheorem{problem}{Problem}
\newtheorem{rem}{Remark}
\newtheorem{prop}{Proposition}

\begin{document}
	\title{Nonlinear Cooperative  Output Regulation with Input Delay Compensation}
	\author{Shiqi Zheng, \IEEEmembership{Senior Member, IEEE}, Choon Ki Ahn, \IEEEmembership{Senior Member, IEEE}, 
		Xiaowei Jiang, \IEEEmembership{Senior Member, IEEE},  \\  Huaicheng Yan, \IEEEmembership{Senior Member, IEEE},  and Peng Shi, \IEEEmembership{Fellow, IEEE}
		\thanks{Shiqi Zheng and Xiaowei Jiang are with the School of Automation, China University of
			Geosciences, Wuhan 430079, China, also with the Hubei Key Laboratory of
			Advanced Control and Intelligent Automation for Complex Systems, Wuhan
			430074, China, also with the Engineering Research Center of Intelligent
			Technology for Geo-Exploration, Ministry of Education, Wuhan 430074,
			China. Shiqi Zheng is also with the Shaoyang Institute of Advanced Manufacturing
			Technology, Shaoyang 422000, China (e-mail: zhengshiqi@cug.edu.cn)}
		\thanks{Choon Ki Ahn is  with the School of Electrical Engineering, Korea University,
			Seoul 136-701, South Korea (e-mail: hironaka@korea.ac.kr).}
		\thanks{Huaicheng Yan is with the Key Laboratory of Smart Manufacturing in
			Energy Chemical Process of Ministry of Education, East China University of
			Science and Technology, Shanghai 200237, China}
	  \thanks{Peng Shi is with the School of Electrical and Mechanical
		Engineering, The University of Adelaide, Adelaide, SA 5005,
		Australia (e-mail: peng.shi@adelaide.edu.au).}
	
}
	
	\maketitle
	
	\begin{abstract}
		This paper investigates the cooperative output regulation (COR)
		of  nonlinear multi-agent systems (MASs) with long input delay based
		on periodic event-triggered mechanism. Compared with other
		mechanisms, periodic event-triggered control can automatically guarantee
		a Zeno-free behavior and avoid the continuous monitoring of triggered
		conditions. First, a new periodic event-triggered distributed observer,
		which is based on the fully asynchronous communication data, is proposed
		to estimate the leader information. Second, a new distributed predictor
		feedback control method is proposed for the considered nonlinear MASs
		with input delay. By coordinate transformation, the MASs are mapped
		into new coupled ODE-PDE target systems with some disturbance-like
		terms. Then, we show that the COR problem is solvable. At last, to
		further save the communication resource, a periodic event-triggered
		mechanism is considered in the sensor-to-controller transmission in every
		agent. A new periodic event-triggered filter is proposed to deal with
		the periodic event-triggered feedback data. The MASs with input
		delay are mapped into coupled ODE-PDE target systems with sampled
		data information. Then, Lyapunov-Krasovskii functions are constructed
		to demonstrate the exponential stability of the MASs. Simulations verify the validity of the proposed results.
	\end{abstract}
	
	
	\section{Introduction}
	
	Recently, the cooperative output regulation (COR) problem of
	multi-agent systems (MASs) has drawn increasing interest \cite{key-1,key-2-2,key-2-2a,key-2-2aa}.
	The COR problem is an extension from the output regulation of a single
	system to MASs. It aims to regulate the output of each follower in
	MASs to a given reference signal/leader while simultaneously rejecting
	the external disturbance. The main difference between the COR problem
	and the classical output regulation problem is that the controller should
	be designed in a distributed way. That is each follower can only rely on
	the information of its neighbors. COR has wide applications, such
	as the formation of multi-vehicle and consensus of multiple robot manipulators
	$etc$. So far, there is a lot of excellent work on the COR problem,
	see the pioneering works in \cite{key-3-1,key-4-1,key-4a}.
	
	In order to reduce the communication burden and save energy for MASs,
	event-triggered cooperative control strategies have gained increasing
	attention. In event-triggered control, the violation of some designed
	conditions will trigger a data-transmission event and then the control
	effort will be updated based on the newly transmitted data. In this
	way, traditional continuous communication will be avoided, thus
	a lot of communication/energy resources can be saved. Various kinds
	of event-triggered control strategies, such as dynamic event-triggered
	control, and self-triggered control $etc$., have been presented \cite{key-5-1,key-6-1,key-6a,key-6aa,key-6aaa}.
	Among them, the periodic event-triggered control (PETC) method has
	attracted a lot of interest \cite{key-7,key-8,key-8a,key-8aa}. Different
	from other event-triggered strategies, the periodic event-triggered method
	only uses the periodic sampled data to design the event-triggered
	condition. This feature can automatically guarantee a Zeno-free behavior
	and avoid the continuous monitoring of triggered conditions. Moreover,
	in most real engineering situations, data transmission is often conducted
	periodically through the network. Therefore, PETC is more suitable
	for real practical network control systems.
	
	Input delay often occurs in MASs. A long input delay is easy to cause
	a loss of stability \cite{key-9-1,key-9a}. Hence, much effort has
	been devoted to dealing with the time-delay MASs. For example, in \cite{key-9-1}
	the consensus conditions were presented for MASs with both time delay
	and measurement noises. \cite{key-10-1} adopted the frequency-domain
	method to analyze the delay margin of the MASs. The above methods
	are sometimes called passive methods because they mainly focus on
	computing a maximal allowable delay that guarantees the stability
	of MASs. However, when the delay is large, one may need to compensate
	for the delay actively to achieve a good control performance. The
	predictor feedback method is one of the most popular methods to handle
	the large delay. In \cite{key-11,key-12-1}, a truncated predictor
	feedback approach was proposed to solve the consensus problem in the
	presence of both communication and input delay. However, as stated
	in \cite{key-13-1,key-14} the method could be sensitive to the delay
	mismatch. In \cite{key-13-1,key-14}, the authors transformed the
	systems with long input delay into coupled ODE-PDE (Ordinary Differential
	Equations-Partial Differential Equations) systems. Then, backstepping
	techniques were employed to analyze the stability \cite{key-15,key-15a}.
	This kind of method can tolerate uncertainties both in the delay
	and system parameters. However\textit{,} \textit{to the best of our
		knowledge, there exists no work about the periodic event-triggered
		COR problem for multi-agent systems with  long input
		delay.}
	
	Therefore, this paper will concentrate on the
	problem of periodic event-triggered COR for a class of nonlinear MASs
	with input delay. The main challenges are as follows:
	
	1) Different from \cite{key-8,key-17}, the network of MASs
	is assumed to be fully asynchronous. Namely, agents have
	various sampling periods and event-triggered time instants (see
	Fig. \ref{fig:1-4}). This assumption is more practical since each
	agent pair may have different communication distances and time clocks.
	However, this assumption makes the analysis more complex;
	
	2) The control law in MASs can only use the information from its neighbors
	and not all the followers can access the information of the leader.
	This makes the predictor feedback technique for a single system invalid;
	
	3) Every follower is described by a nonlinear system with long input
	delay. The considered nonlinear systems can be a high-order/strict-feedback
	nonlinear system. Because we cannot obtain an explicit solution to
	the nonlinear systems, the existing methods \cite{key-13-1,key-14,key-15}
	for linear MASs cannot be directly extended to our case. It further
	complicates the controller design procedure; and
	
	4) The periodic event-triggered data transmission is considered both
	in the network among various agents and the sensor-to-controller
	transmission in every agent. This will bring more difficulties to the problem since the feedback data are all transmitted in a discrete manner.
	
	To tackle the above issues, we present our main contributions:
	\begin{itemize}
		\item A new periodic event-triggered adaptive distributed observer, which
		is based on the fully asynchronous communication data, is proposed
		to estimate the leader information. According to the stability analysis
		of systems with multiple time delays, we can prove that the estimation error
		can converge to the origin exponentially. Moreover, we present a relationship
		between the maximally allowable sampling periods and the observer
		gains.
		\item A new distributed predictor feedback control method is proposed for
		a class of nonlinear MASs with long input delay. Based on the certainty
		equivalence principle and state transformation technique, the MASs
		are mapped into new coupled ODE-PDE target systems with some disturbance-like
		terms. Then, we show that the COR problem is solvable. Moreover, we
		extend our proposed method to strict-feedback  high-order nonlinear
		systems.
		\item In order to further save the communication resource, a periodic event-triggered
		mechanism (PETM) is also considered in the sensor-to-controller channel
		in each follower. A new periodic event-triggered filter is proposed
		to deal with the periodic event-triggered feedback data. By this filter,
		the discrete feedback data are transformed into continuous signals.
		Then, the MASs with input delay are mapped into coupled ODE-PDE target
		systems with sampled data information. This facilitates the subsequent
		controller design.
	\end{itemize}
	\textit{Notations}. Define set $\mathbb{E}(p)$ with respect to a
	non-negative real constant $p$ and a function $f(t):\mathbb{R}\rightarrow\mathbb{R}$.
	If $f(t)\in\mathbb{E}(p)$, then $|f(t)|$ will converge to a set
	containing the origin exponentially, $i.e.$, $|f(t)|\leq c_{1}\mathrm{e}^{-c_{2}t}+\iota(p)$
	where $c_{1},c_{2}$ are two positive constants and $\iota(p):\mathbb{R}\rightarrow\mathbb{R}$
	is a non-decreasing function with respect to $p$ and $\iota(0)=0.$
	
	\section{Problem formulation}
	
	The leader model of the considered MASs is:
	\begin{align}
	\dot{v} & =Sv,\label{eq:1}\\
	y_{0} & =Fv,\label{eq:2-2}
	\end{align}
	where $v\in\mathbb{R}^{n_{v}}$ denotes the reference signal and/or
	external disturbance. $S$ represents the system matrix, which is
	only known by a small portion of followers. $y_{0}\in\mathbb{R}$
	is the output of the leader. $F=[1,0,...,0]$. We assume that the
	real parts of the eigenvalues of the system matrix $S$ are zero.
	This implies that the leader system is marginally stable. By selecting
	an appropriate matrix $S$, various
	kinds of bounded signals, including step, cosine $etc$, can be produced. 
	
	The followers are expressed as:
	\begin{align}
	\dot{X}_{i} & =f_{i}(X_{i})+U_{i}(t-D_{i}),\label{eq:6-2}\\
	Y_{i} & =X_{i},\label{eq:7-2}\\
	e_{i} & =Y_{i}-y_{0}, i=1,2,...,N, \label{eq:8-4}
	\end{align}
	where $X_{i}\in\mathbb{R}$, $U_{i}\in\mathbb{R}$, $Y_{i}\in\mathbb{R}$,
	$e_{i}\in\mathbb{R}$ represent the system state, control input, system
	output and regulation error respectively. \textcolor{black}{$D_{i}>0$ denotes the long
	input delay, which is a positive constant, and can be different for every follower. }
	By long delay, we mean
	that the delay \textit{$D_{i}$} can be any large positive constant.
	$f_{i}(X_{i}):\mathbb{R}\rightarrow\mathbb{R}$ is a known smooth
	nonlinear function satisfying $\left|\frac{df_{i}(X_{i})}{dX_{i}}\right|\leq\ell$
	where $\ell$ is a positive constant. 
	
	We use a graph $\mathcal{G}$  to represent the communication
	network of the MASs consisting of $1$ leader and $N$ followers. See \cite{key-4-1} for more information about the definition of graph. \textcolor{black}{$\mathcal{G}$ is assumed to be directed and at least have a directed spanning tree.}
%
%

	According to the above MASs, the periodic event-triggered COR problem
	is:
	\begin{problem}
		\label{prob:Given-a-multi-agent}Consider the MASs described by (\ref{eq:1})-(\ref{eq:8-4})
		together with the  graph $\mathcal{G}$. The purpose is to design
		a PETC strategy for every follower
		such that $\underset{t\rightarrow+\infty}{\lim}|e_{i}(t)|=0$ where
		$i\in\{1,2,...,N\}$.
	\end{problem}

    \begin{rem}
    \textcolor{black}{It is noted that  here we only consider first-order nonlinear system, however, the proposed  method can be easily extended to high-order or strict-feedback nonlinear systems. Please see Appendix F in []. Meanwhile, the presented nonlinear model  is quite general and can describe many real phenomena in real practice, such as  formation control of a group of unmanned automobiles or unmanned aerial vehicles, and cooperative control of robot manipulators or servo motor drive systems.}
    \end{rem}

%
%
	
	\section{Periodic event-triggered adaptive distributed observer}
	
\begin{figure}
		\begin{centering}
			\includegraphics[scale=0.7]{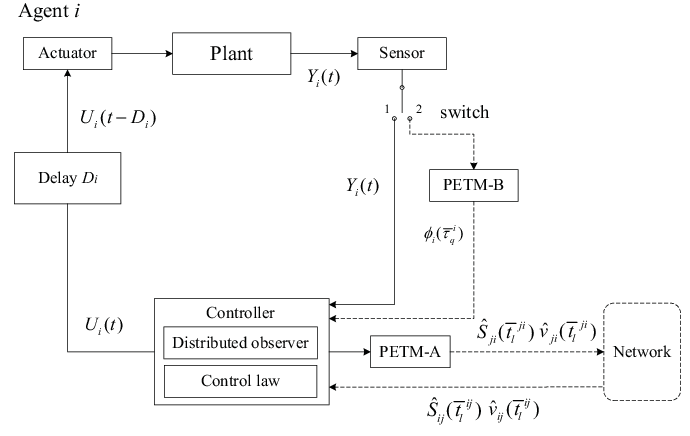}
			\par\end{centering}
		\caption{\label{fig:1-3}Proposed control scheme. PETM-A and PETM-B are two
			periodic event-triggered mechanisms given in (\ref{eq:26}) and (\ref{eq:26-1-1-2-1})
			respectively. PETM-A denotes the PETM between each agent. PETM-B denotes
			the PETM in the sensor-to-controller transmission for every agent, which
			will be explained in Section V.}
	\end{figure}

	Fig. \ref{fig:1-3} shows the presented PETC scheme. It includes a distributed
	observer and a control law. Each agent implements a 
	distributed observer to estimate the dynamics of the leader ($S$
	and $v$) based on  its neighbors. The PETM determines
	whether or not each agent will broadcast its data to its neighbors.
	The control law utilizes the estimated values from the periodic event-triggered
	distributed observer and the feedback data from the sensor to produce
	an appropriate control effort. The control effort will be sent to
	the actuator side which suffers from a long delay $D_{i}$. In this
	section, we will present the periodic event-triggered distributed
	observer for the considered MASs. In Section IV, we will explain the
	proposed control law in detail.
	
\begin{figure}
		\begin{centering}
			\includegraphics[scale=0.8]{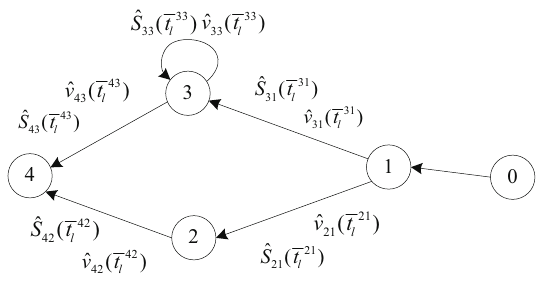}
			\par\end{centering}
		\caption{\label{fig:1-4}Fully asynchronous communication. Each agent pair
			has different sampling periods and event-triggered time instants.}
	\end{figure}

	As illustrated in Fig. \ref{fig:1-4}, for each agent pair $(j,i)$,
	let $\Xi_{T}^{ij}=\{t_{0}^{ij},t_{1}^{ij},...,t_{k}^{ij},...\}(i,j\in\{1,2,...,N\})$
	with $0=t_{0}^{ij}<t_{1}^{ij}<\cdots<t_{k}^{ij}<\cdots$ denote the
	asynchronous sampling time instants such that $t_{k}^{ij}\triangleq kT^{ij}$
	with sampling period $T^{ij}$. \textcolor{black}{If $i\neq j, t_{k}^{ij}$ means the sampling time instants between agent $i$ and agent $j$. If $i=j, t_{k}^{ii}$ represents the sampling time instants for agent $i$ itself, that is the agent samples its own states. }
	For simplicity, $l$ denotes the latest index of the sampling time instants
	for each agent pair. Then,  the following  distributed observer is presented for every follower
	\begin{align}
	\dot{\hat{S}}_{i} & =\kappa_{1}\sum_{j=0}^{N}a_{ij}(\hat{S}_{j}(\overline{t}_{l}^{ij})-\hat{S}_{i}(\overline{t}_{l}^{ii})),\label{eq:5-1-1-1}\\
	\dot{\hat{v}}_{i} & =\hat{S}_{i}\hat{v}_{i}+\kappa_{2}\sum_{j=0}^{N}a_{ij}(\overline{v}_{j}(t,\overline{t}_{l}^{ij})-\overline{v}_{i}(t,\overline{t}_{l}^{ii})),\label{eq:6-1-1}
	\end{align}
	where $i=1,2,...,N$, $\hat{S}_{0}(t)\equiv S$, $\kappa_{1},\kappa_{2}>0$
	are two constants. $\hat{v}_{i}$ and $\hat{S}_{i}$ are the estimations
	of the real values $v_{i}$ and $S_{i}$ for each agent.
	\[
	\overline{v}_{j}(t,\overline{t}_{l}^{ij})=\mathrm{e}^{\hat{S}_{j}(\overline{t}_{l}^{ij})(t-\overline{t}_{l}^{ij})}\hat{v}_{j}(\overline{t}_{l}^{ij}),
	\]
	Specifically, if $i=j$,
	\begin{equation}
	\overline{v}_{i}(t,\overline{t}_{l}^{ii})=\mathrm{e}^{\hat{S}_{i}(\overline{t}_{l}^{ii})(t-\overline{t}_{l}^{ii})}\hat{v}_{i}(\overline{t}_{l}^{ii}).\label{eq:8}
	\end{equation}
	If $j=0$,
	\[
	\overline{v}_{0}(t,\overline{t}_{l}^{i0})=\mathrm{e}^{S(t-t_{l}^{i0})}\hat{v}_{0}(t_{l}^{i0})=v(t)
	\]
	with $\hat{v}_{0}(t)\triangleq v(t)$.
	
	Agent $j$ will send data to agent $i$ at time instant $\overline{t}_{l}^{ij}$,
	which denotes the latest event-triggered time instant. Specifically,
	the follower $j$ will transmit $\hat{S}_{i}(\overline{t}_{l}^{ij})$
	and $\hat{v}_{i}(\overline{t}_{l}^{ij})$ to agent $i$ on $\overline{t}_{l}^{ij}$.
	$\overline{t}_{l}^{ij}$ is determined by:
	\begin{equation}
	\overline{t}_{l+1}^{ij}=\mathrm{inf}\{\tau>\overline{t}_{l}^{ij}|\tau\in\Xi_{T}^{ij},h_{S}^{ij}(\cdot)>0,\mathrm{or}\thinspace h_{v}^{ij}(\cdot)>0\},\label{eq:26}
	\end{equation}
	where
	\[
	h_{S}^{ij}(\cdot)=h_{S}^{ij}(\tau,\overline{t}_{l}^{ij})=||\hat{S}_{i}(\tau)-\hat{S}_{i}(\overline{t}_{l}^{ij})||-\delta_{S}\mathrm{e}^{-\gamma_{S}\tau},
	\]
	\[
	h_{v}^{ij}(\cdot)=h_{v}^{ij}(\tau,\overline{t}_{l}^{ij})=||\hat{v}_{i}(\tau)-\overline{v}_{i}(\tau,\overline{t}_{l}^{ij})||-\delta_{v}\mathrm{e}^{-\gamma_{v}\tau}
	\]
	with $\delta_{S},\delta_{v},\gamma_{S},\gamma_{v}>0$. $l$
	denotes the latest index of the periodic event-triggered instant
	for every agent pair with some abuse of notation. \textcolor{black}{When
		$h_{S}^{ij}(\cdot)>0$ or $h_{v}^{ij}(\cdot)>0$, the event-triggered
		condition is satisfied and $\hat{S}_{i},\hat{v}_{i}$ will be transmitted.
		Then, the current time $t$ will be set to $\overline{t}_{l}^{ij}$.
		This implies that $h_{S}^{ij}(\cdot)=0<\delta_{S}\mathrm{e}^{-\gamma_{S}\tau},h_{v}^{ij}(\cdot)=0<\delta_{v}\mathrm{e}^{-\gamma_{v}\tau}$.
		Therefore, the event-triggered condition will not be triggered consistently. }
	
	Then, we have:
	\begin{thm}
		\label{thm:1} Consider the MASs (\ref{eq:1})-(\ref{eq:8-4}) with
		the distributed observer (\ref{eq:5-1-1-1})-(\ref{eq:6-1-1}),
		there exists a positive constant $M$ such that for any $\kappa T\leq M$
		with $\kappa\triangleq\max\{\kappa_{1},\kappa_{2}\}$ and $T\triangleq\underset{i,j\in\{1,2,...,N\}}{\max}\{T^{ij}\}$,
		the estimation errors $\tilde{S}_{i}\triangleq\hat{S}_{i}-S$ and
		$\tilde{v}_{i}\triangleq\hat{v}_{i}-v(i=1,...,N)$ will tend to
		the origin exponentially.
	\end{thm}
	\begin{proof}
		See  Appendix A.
	\end{proof}
    \begin{rem}
    \textcolor{black}{It is noted that we aim to present a distributed control method for the MASs. Meanwhile, we have assumed that only a small portion of the followers can have access to the leader. That is only those followers that are directly connected to the leader can access the leader. Therefore, in order to design a distributed controller in this situation, we have designed a distributed observer for each follower. Based on this observer, every follower can estimate the leader states only based on its neighborhood information. Then, using the estimated leader information, we can design distributed controller for every follower. }
    	
   \textcolor{black}{ Note that the $N$ times estimations in (\ref{eq:5-1-1-1})-(\ref{eq:6-1-1}) mean that every follower needs a distributed observer to obtain the leader information, especially those followers that cannot obtain the leader states directly. Though there are N observers for the multi-agent systems,  we only need to design one observer for each agent. Therefore, the computation complexity does not increase too much for each agent. In fact, we only need to implement two extra simply ODEs (\ref{eq:5-1-1-1})-(\ref{eq:6-1-1}) for each agent. }
    \end{rem}
     
	\begin{rem}
	\textcolor{black}{	Different from the existing works \cite{key-8,key-17}, the proposed
		distributed observer can deal with fully asynchronous periodic event-triggered
		transmission data, $i.e.,$ every agent pair has a different sampling
		period. Nevertheless, in \cite{key-8,key-17} each agent has the same
		sampling period for communication with its neighbors. The fully asynchronous
		periodic event-triggered transmission data is more practical in real
		engineering applications due to each agent pair may have different
		communication distances and time clocks. 			
		Moreover, from the above result, we can see that
		to guarantee the effectiveness of the proposed observer, the observer parameters should satisfy the inequality
		$\kappa T\leq M$. This implies that the maximally allowable sampling
		periods $T^{ij}$ can be made large by decreasing the observer gain
		$\kappa$ with a sacrifice of convergence rate. A possible choice
		of $M$ can be found in Appendix A. If the observer gains and sampling period are too large, we will 
		lose observability. In these situations, we can decrease the observer gain and the sampling period to remedy the problem. }
%
	\end{rem}

	\section{Control law }
	
	
	\subsection{Design of the control law}
	
	In order to handle the long input delay, future states will be
	predicted. Then, the predicted states will be used to design the controller.
	
	Let
	\begin{equation}
	\overline{X}_{i}=e_{i}=X_{i}-Fv.\label{eq:37-1}
	\end{equation}
	Then, the system (\ref{eq:6-2})-(\ref{eq:8-4}) becomes
	\begin{align}
	\dot{\overline{X}}_{i}= & f_{i}(\overline{X}_{i}+Fv)-f_{i}(Fv)\nonumber \\
	& +f_{i}(Fv)-FSv+U_{i}(t-D_{i})\nonumber \\
	= & \overline{f}_{i}(\overline{X}_{i},v)+(u^{i}(0,t)-FSv+f_{i}(Fv))\nonumber \\
	= & \overline{f}_{i}(\overline{X}_{i},v)+\overline{u}^{i}(0,t),\label{eq:34}\\
	e_{i}= & \overline{X}_{i},
	\end{align}
	where $\overline{f}_{i}(\overline{X}_{i},v)=f_{i}(\overline{X}_{i}+Fv)-f_{i}(Fv)$,
	$R_{i}(v,S)\triangleq FSv-f_{i}(Fv)$.
	\begin{align}
	u^{i}(x,t) & \triangleq U_{i}(t+(x-1)D_{i}),\label{eq:38-1}\\
	\overline{u}^{i}(x,t) & \triangleq u^{i}(x,t)-R_{i}(\mathrm{e}^{SD_{i}x}v,S),x\in[0,1]\label{eq:39-1}
	\end{align}
	represent the distributed inputs.

	According to the above dynamics, we can predict the state. Let
	\begin{align*}
	\chi^{i}(x,t) & \triangleq\overline{X}_{i}(t+D_{i}x)\\
	& =\overline{X}_{i}(t)+D_{i}\int_{0}^{x}\left(\overline{f}_{i}(\chi^{i}(y,t),\mathrm{e}^{SD_{i}y}v)+\overline{u}^{i}(y,t)\right)dy.
	\end{align*}
	Then, the input $\overline{u}^{i}(1,t)$ is given by:
	\begin{align*}
	& \overline{u}^{i}(1,t)\\
	= & U_{i}-R_{i}(\mathrm{e}^{SD_{i}}v,S)\\
	\triangleq & K_{i}\overline{X}_{i}(t+D_{i})\\
	= & K_{i}\chi^{i}(0,t)+K_{i}D_{i}\int_{0}^{1}\left(\overline{f}_{i}(\chi^{i}(y,t),\mathrm{e}^{SD_{i}y}v)+\overline{u}_{i}(y,t)\right)dy,
	\end{align*}
	where $K_{i}$ is the control gain. The above controller is in fact
	a proportional state feedback control law. From (\ref{eq:39-1}),
	the controller $u^{i}(1,t)=U_{i}(t)$ is given by
	\begin{align}
	& u^{i}(1,t)=U_{i}(t)\nonumber \\
	= & K_{i}\overline{X}_{i}(t+D_{i})+R_{i}(\mathrm{e}^{SD_{i}}v,S)\nonumber \\
	= & K_{i}D_{i}\int_{0}^{1}\left(\overline{f}_{i}(\chi_{i}(y,t),\mathrm{e}^{SD_{i}y}v)+\overline{u}_{i}(y,t)\right)dy\nonumber \\
	& +K_{i}\overline{X}_{i}(t)+R_{i}(\mathrm{e}^{SD_{i}}v,S).\label{eq:35}
	\end{align}
	Based on the leader dynamics (\ref{eq:1}) and (\ref{eq:39-1}), it
	follows that
	\begin{align*}
	\overline{u}^{i}(0,t) & =u^{i}(0,t)-R_{i}(v,S)\\
	& =U_{i}(t-D_{i})-R_{i}(v,S)\\
	& =K_{i}\overline{X}_{i}(t)+R_{i}(\mathrm{e}^{SD_{i}}v(t-D_{i}),S)-R_{i}(v,S)\\
	& =K_{i}\overline{X}_{i}(t).
	\end{align*}
	This implies that (\ref{eq:34}) becomes
	\begin{align*}
	\dot{\overline{X}}_{i} & =\overline{f}_{i}(\overline{X}_{i},v)+K_{i}\overline{X}_{i}.
	\end{align*}
	Since $\left|\frac{df_{i}(X_{i})}{dX_{i}}\right|\leq\ell$, we have
	$|\overline{f}_{i}(\overline{X}_{i},v)|\leq\ell|\overline{X}_{i}|$
	by Mean Value Theorem. Then, we can choose a control gain $K_{i}$
	such that $\ell+K_{i}<0$. Then, by selecting the Lyapunov function
	$V_{i}=\frac{1}{2}\overline{X}_{i}^{2}$, we can  verify that
	$\overline{X}_{i}(t)\rightarrow0$ as $t\rightarrow+\infty$. By (\ref{eq:8-4}),
	$e_{i}(t)$ will converge to zero. Summarizing the above analysis,
	we have:
	\begin{prop}
		\label{prop:Consider-the-multi-agent-1}Given the MASs consisting
		of the leader (\ref{eq:1}) and nonlinear followers (\ref{eq:6-2})-(\ref{eq:8-4}),
		the controller in the form of (\ref{eq:35}) can solve Problem \ref{prob:Given-a-multi-agent}.
	\end{prop}
	
	Note that though the above result is true, the controller (\ref{eq:35})
	does not satisfy our design requirement. The reasons are as follows:
	First, the matrices $S$ and the state $v$ are from the leader. They
	are not known by all the followers; Second, the long input delay $D_{i}$
	may not be measured accurately. Next, we will re-design the controller.
	Inspired by the certainty equivalence principle, according to (\ref{eq:38-1})-(\ref{eq:39-1}),
	let
	\begin{equation}
	\check{u}^{i}(x,t)=U_{i}(t+(x-1)\hat{D}_{i}),\label{eq:22-1-1}
	\end{equation}
	\begin{equation}
	\hat{u}^{i}(x,t)=\check{u}^{i}(x,t)-R_{i}(\mathrm{e}^{\hat{S}_{i}\hat{D}_{i}x}\hat{v}_{i},\hat{S}_{i}),\label{eq:23-1-1}
	\end{equation}
	\begin{align}
	\hat{u}^{i}(1,t) & \triangleq K_{i}\hat{\chi}^{i}(0,t)\nonumber \\
	& +K_{i}D_{i}\int_{0}^{1}\left(\overline{f}_{i}(\hat{\chi}^{i}(y,t),\mathrm{e}^{SD_{i}y}v)+\overline{u}^{i}(y,t)\right)dy,
	\end{align}
	where
	\begin{align}
	& \hat{\chi}^{i}(x,t)\nonumber \\
	= & \hat{X}_{i}(t)+\hat{D}_{i}\int_{0}^{x}\left(\overline{f}_{i}(\hat{\chi}^{i}(y,t),\mathrm{e}^{\hat{S}_{i}\hat{D}_{i}y}\hat{S}_{i}\hat{v}_{i})+\hat{u}^{i}(y,t)\right)dy.\label{eq:17}
	\end{align}
	The output of the controller $U_{i}(t)$ is designed as
	\begin{align}
	& U_{i}(t)=\check{u}^{i}(1,t)\nonumber \\
	= & K_{i}\hat{\chi}^{i}(0,t)\nonumber \\
	& +K_{i}\hat{D}_{i}\int_{0}^{1}\left(\overline{f}_{i}(\hat{\chi}^{i}(y,t),\mathrm{e}^{\hat{S}_{i}\hat{D}_{i}y}\hat{S}_{i}\hat{v}_{i})+\hat{u}^{i}(y,t)\right)dy\nonumber \\
	& +R_{i}(\mathrm{e}^{\hat{S}_{i}\hat{D}_{i}}\hat{v}_{i},\hat{S}_{i}),\label{eq:45-1}
	\end{align}
	where we use $\hat{D}_{i}$ to denote the estimation of the real value
	$D_{i}$, which indicates that there may exist a delay mismatch. $\hat{D}_{i}$
	can be obtained by engineering experience. Let $\widetilde{D}_{i}=D_{i}-\hat{D}_{i}$
	denote the delay mismatch. $\hat{S}_{i},\hat{v}_{i}$ are obtained
	from the periodic event-triggered distributed observer in Section
	III. $\hat{X}_{i}=X_{i}-F_{i}\hat{v}_{i}$ is an estimation of $\overline{X}_{i}$
	in (\ref{eq:37-1}).
	
	According to the above analysis, we have
	\begin{thm}
		\label{thm:1-1-1-1}Given the MASs with leader (\ref{eq:1})-(\ref{eq:2-2})
		and nonlinear followers (\ref{eq:6-2})-(\ref{eq:8-4}). There exists
		a control law (\ref{eq:45-1}) with a
		distributed observer (\ref{eq:5-1-1-1})-(\ref{eq:6-1-1}) such that
		Problem 1 is solvable. Specifically, there exists a sufficiently small
		positive constant $D^{*}$ such that for $\forall|\hat{D}_{i}-D_{i}|\leq D^{*}(i=1,2,..,N)$,
		the  error $|e_{i}(t)|\in\mathrm{\mathbb{E}}(\widetilde{D}_{i})$.
	\end{thm}
	\begin{proof}
		We put the proof in Section IV-B.
	\end{proof}
	\begin{rem}
		The above result indicates the output regulation error will converge
		to a set around the origin. The neighborhood is determined by the
		delay mismatch $\widetilde{D}_{i}$. In fact, a better estimation
		of the input delay $D_{i}$ can result in more accurate control performance.
		If there exists no delay mismatch, then the output regulation will
		converge to the origin exponentially.
	\end{rem}
	
	\subsection{Proof of Theorem \ref{thm:1-1-1-1}}
	
	Given the following coordinate transformation
	\begin{align}
	\hat{w}^{i}(x,t) & \triangleq\hat{u}^{i}(x,t)-K_{i}\hat{\chi}^{i}(x,t),\label{eq:47}
	\end{align}
	we have:
	\begin{prop}
		\label{prop:Under-the-backstepping}Under the coordinate transformation
		(\ref{eq:47}), the system (\ref{eq:6-2})-(\ref{eq:8-4}) is mapped
		into the following target system:
		
		$\overline{X}$-dynamics:
		\begin{align}
		\dot{\overline{X}}_{i}= & \overline{f}_{i}(\overline{X}_{i},v)+K_{i}\overline{X}_{i}+\widetilde{u}^{i}(0,t)+\hat{w}^{i}(0,t)+\zeta_{i0}(t).\label{eq:19}
		\end{align}
		
		$u$-dynamics:
		\begin{align}
		D_{i}\widetilde{u}_{t}^{i}(x,t) & =\widetilde{u}_{x}^{i}(x,t)-\frac{\widetilde{D}_{i}}{\hat{D}_{i}}\hat{u}_{x}^{i}(x,t)+\zeta_{i1}(x,t),\label{eq:21}\\
		\widetilde{u}^{i}(1,t) & =R_{i}(\mathrm{e}^{\hat{S}_{i}\hat{D}_{i}}\hat{v}_{i},\hat{S}_{i})-R_{i}(\mathrm{e}^{SD_{i}}v,S).\label{eq:22}
		\end{align}
		
		$w$-dynamics:
		\begin{align}
		\hat{D}_{i}\hat{w}_{t}^{i}(x,t)= & \hat{w}_{x}^{i}(x,t)+\hat{D}_{i}\widetilde{u}^{i}(0,t)\Theta^{i}(x,t)+\zeta_{i2}(x,t),\label{eq:23}\\
		\hat{w}^{i}(1,t)= & 0.\label{eq:24}
		\end{align}
		
		$w_{x}$-dynamics:
		\begin{align}
		\hat{D}_{i}\hat{w}_{xt}^{i}(x,t)= & \hat{w}_{xx}^{i}(x,t)+\hat{D}_{i}\widetilde{u}^{i}(0,t)\Theta_{x}^{i}(x,t)+\zeta_{i3}(x,t),\label{eq:25}\\
		\hat{w}_{x}^{i}(1,t)= & -\hat{D}_{i}\widetilde{u}^{i}(0,t)\Theta^{i}(1,t)-\zeta_{i2}(1,t),\label{eq:26-1}
		\end{align}
		where $i=1,2,...,N$, $\widetilde{u}^{i}(x,t)=\overline{u}^{i}(x,t)-\hat{u}^{i}(x,t)$,
		$\zeta_{i0}(t),\zeta_{ij}(x,t)(j=1,2,3)\in\mathrm{\mathbb{E}}(0)$
		for $\forall x\in[0,1]$, and
		\[
		\Theta^{i}(x,t)=\mathrm{e}^{\intop_{0}^{x}\hat{D}_{i}\frac{\partial\overline{f}_{i}}{\partial\hat{\chi}^{i}}(y,t)dy}.
		\]
	\end{prop}
	\begin{proof}
		See Appendix B.
	\end{proof}
	\begin{rem}
		Because we cannot obtain an explicit solution to the nonlinear system
		(\ref{eq:6-2}), the proposed mapped target system is more complicated
		than the linear system \cite{key-13-1,key-14}. Moreover, since only
		part of the followers can access the
		leader, the target system contains some disturbance-like terms $\zeta_{i0}(t),\zeta_{ij}(x,t)(j=1,2,3)$.
	\end{rem}
	
	Next, let us consider the following Lyapunov function
	\begin{align}
	V_{i}= & \sum_{i=1}^{4}\lambda_{j}V_{ij},\label{eq:35-1}
	\end{align}
	where $\lambda_{1}=1$, $\lambda_{j}(j=2,3,4)$ are positive parameters.
	$V_{i1}=\overline{X}_{i}^{2}/2$, $V_{i2}=\frac{D_{i}}{2}\int_{0}^{1}(1+x)\widetilde{u}^{i}(x,t)^{2}dx$,
	$V_{i3}=\frac{\hat{D}_{i}}{2}\int_{0}^{1}(1+x)\hat{w}^{i}(x,t)^{2}dx$,
	$V_{i4}=\frac{\hat{D}_{i}}{2}\int_{0}^{1}(1+x)\hat{w}_{x}^{i}(x,t)^{2}dx$.
	
	Then, we have:
	\begin{prop}
		\label{prop:There-exist-positive}There exists a positive constant
		$D^{*}$ and small enough $\lambda_{j}(j=2,3,4)$ such that for $\forall|\widetilde{D}_{i}|\leq D^{*},$
		\begin{equation}
		\dot{V}_{i}\leq-aV_{i}+\varepsilon_{i}(t),\label{eq:32}
		\end{equation}
		where $i=1,2,...,N$, $a>0$ is a  constant, $\varepsilon_{i}(t)\in\mathrm{\mathbb{E}}(\widetilde{D}_{i})$.
	\end{prop}
	\begin{proof}
	See Appendix C.
	\end{proof}
	
	By solving the above inequality (\ref{eq:32}) for $V_{i}$, we can
	complete the proof of Theorem \ref{thm:1-1-1-1}.
	
	\section{Periodic event-triggered control law}
	
     We give an extension to the main results. In order to further reduce the data transmission, a PETM is also utilized in
	the sensor-to-controller channel (the switch
	is on node 2 in Fig. \ref{fig:1-3}).
	
	\subsection{Controller design}
	
	For simplicity, it is assumed that there is no delay mismatch, $i.e.$,
	$D_{i}=\hat{D}_{i}$. In this case, the control law is also given
	by (\ref{eq:37-1}) except that $\hat{X}_{i}$ is determined by the
	following new periodic event-triggered filter.
	\begin{align}
	\dot{\hat{X}}_{i} & =\overline{f}_{i}(\hat{X}_{i},\hat{v}_{i})+\hat{u}^{i}(0,t)-L_{i}(\phi_{i}(\overline{\tau}_{q}^{i})-\hat{X}_{i}(\tau_{p}^{i})),\label{eq:45-2}\\
	\phi_{i}(t) & =Y_{i}(t)-F_{i}\overline{v}_{i}(t,\overline{t}_{l}^{ii}),\thinspace t\in[\tau_{p}^{i},\tau_{p+1}^{i}),\label{eq:23-2}
	\end{align}
	where $L_{i}$ is a design parameter. $0=\tau_{0}^{i}<\tau_{1}^{i}<\cdots<\tau_{p}^{i}<\cdots$
	represent the sampling time instants with the sampling period $\mathcal{T}_{i}=\tau_{p+1}^{i}-\tau_{p}^{i}$.
	$0=\overline{\tau}_{0}^{i}<\overline{\tau}_{1}^{i}<\cdots<\overline{\tau}_{q}^{i}<\cdots$
	are the periodic event-triggered instants. On  $\overline{\tau}_{q}^{i}$,
	the sensor will send $\phi_{i}(\overline{\tau}_{q}^{i})$ to the
	controller. $\overline{\tau}_{q+1}^{i}$ is given by:
	\begin{equation}
	\overline{\tau}_{q+1}^{i}=\mathrm{inf}\{\tau>\overline{\tau}_{q}^{i}|\tau\in\Xi_{\mathcal{T}}^{i},h_{\phi}^{i}(\tau,\overline{\tau}_{q}^{i})>0\},\label{eq:26-1-1-2-1}
	\end{equation}
	where $\Xi_{\mathcal{T}}^{i}=\{\tau_{0}^{i},\tau_{1}^{i},...,\tau_{p}^{i},...\},$
	\begin{equation}
	h_{\phi}^{i}(\tau,\overline{\tau}_{q}^{i})=||\phi_{i}(\tau)-\phi_{i}(\overline{\tau}_{q}^{i})||-\delta_{\phi}\mathrm{e}^{-\gamma_{\phi}\tau}\label{eq:17-2-1-1}
	\end{equation}
	with positive constants $\delta_{\phi},\gamma_{\phi}$. Since
	the followers only need to measure their own outputs $Y_{i}(t)$ continuously,
	the value $\overline{v}_{i}(t,\overline{t}_{l}^{ii})$ is determined
	by the discrete measurements $\hat{S}_{i}(\overline{t}_{l}^{ii}),\hat{v}_{i}(\overline{t}_{l}^{ii})$
	in the distributed observer (\ref{eq:8}).
	
	Then, we have:
	\begin{thm}
		\label{thm:1-1-1-1-1}Given the MASs with leader (\ref{eq:1})-(\ref{eq:2-2})
		and nonlinear followers (\ref{eq:6-2})-(\ref{eq:8-4}). There exists
		a control law (\ref{eq:45-1}) with  a distributed
		observer (\ref{eq:5-1-1-1})-(\ref{eq:6-1-1}) and a
		filter (\ref{eq:45-2})-(\ref{eq:23-2}) such that Problem 1 is solvable
		and the output regulation error $|e_{i}(t)|\in\mathrm{\mathbb{E}}(0)$.
	\end{thm}
	\begin{proof}
		See Section V-B.
	\end{proof}
	\begin{rem}
		It is noted that due to the periodic event-triggered feedback data
		are transmitted in a discrete manner, $\phi_{i}(\overline{\tau}_{q}^{i}),\hat{X}_{i}(\tau_{p}^{i})$
		cannot be directly used in the controller design. Hence, we propose
		a periodic event-triggered filter to obtain an estimation of the state
		$\overline{X}_{i}$.
	\end{rem}
	
	\subsection{Proof of Theorem \ref{thm:1-1-1-1-1}}
	
	Similar to Proposition \ref{prop:Under-the-backstepping}, we have:
	\begin{prop}
		\label{prop:Under-the-backstepping-1}Under the transformation (\ref{eq:47}),
		the system (\ref{eq:6-2})-(\ref{eq:8-4}) with the periodic event-triggered
		filter (\ref{eq:45-2})-(\ref{eq:23-2}) is mapped into the following
		target system:
		
		$\widetilde{X}$-dynamics and $\hat{X}$-dynamics:
		\begin{align}
		\dot{\widetilde{X}}_{i}= & \overline{f}_{i}(\overline{X}_{i},v)-\overline{f}_{i}(\hat{X}_{i},\hat{v}_{i})+L_{i}\widetilde{X}_{i}\nonumber \\
		& +L_{i}\widetilde{X}_{i}(\tau_{p}^{i})-L_{i}\widetilde{X}_{i}+\xi_{i0}(t),\label{eq:33-1}
		\end{align}
		\begin{align}
		\dot{\hat{X}}_{i}= & \overline{f}_{i}(\hat{X}_{i},\hat{v}_{i})+K_{i}\hat{X}_{i}+\hat{w}^{i}(0,t)\nonumber \\
		& +K_{i}\hat{X}_{i}(\tau_{p}^{i})-K_{i}\hat{X}_{i}\nonumber \\
		& -L_{i}\widetilde{X}_{i}(\tau_{p}^{i})+\xi_{i0}(t).\label{eq:34-1}
		\end{align}
		
		$w$-dynamics:
		\begin{align}
		\hat{D}_{i}\hat{w}_{t}^{i}(x,t)= & \hat{w}_{x}^{i}(x,t)\nonumber \\
		& +K_{i}\hat{D}_{i}\Theta^{i}(x,t)(K_{i}\hat{X}(\tau_{p}^{i})-K_{i}\hat{X})\nonumber \\
		& +K_{i}\hat{D}_{i}\Theta^{i}(x,t)L_{i}\widetilde{X}_{i}(\tau_{p}^{i})+\xi_{i1}(x,t),\label{eq:35-2}\\
		\hat{w}^{i}(1,t)= & 0,\label{eq:36-1}
		\end{align}
		where $\xi_{i0}(t),\xi_{i1}(x,t)\in\mathrm{\mathbb{E}}(0)$ for $\forall x\in[0,1]$.
	\end{prop}
	\begin{proof}
		See  Appendix D.
	\end{proof}
	\begin{rem}
		Different from \cite{key-13-1,key-14}, the proposed target system
		contains some sampled data information $\hat{X}(\tau_{p}^{i}),\widetilde{X}_{i}(\tau_{p}^{i})$.
		These are used to handle the periodic event-triggered transmission
		data in the sensor-to-controller channel.
	\end{rem}
	
	Next, let us consider the following Lyapunov-Krasovskii function
	\begin{align}
	\mathcal{V}_{i}= & \sum_{i=1}^{5}\lambda_{j}\mathcal{V}_{ij},\label{eq:35-1-1}
	\end{align}
	where $\lambda_{1}=1$, $\lambda_{j}(j=2,...,5)$ are positive parameters.
	$\mathcal{V}_{i1}=\widetilde{X}_{i}^{2}/2$, $\mathcal{V}_{i2}=\hat{X}_{i}^{2}/2$,
	$\mathcal{V}_{i3}=\frac{\hat{D}_{i}}{2}\int_{0}^{1}(1+x)\hat{w}^{i}(x,t)^{2}dx$,
	$\mathcal{V}_{i4}=\int_{t-\mathcal{T}_{i}}^{t}\int_{s}^{t}||\dot{\widetilde{X}}_{i}(\theta)||^{2}d\theta ds$,
	$\mathcal{V}_{i5}=\int_{t-\mathcal{T}_{i}}^{t}\int_{s}^{t}||\dot{\hat{X}}_{i}(\theta)||^{2}d\theta ds$.
	
	Then, we have the following result.
	\begin{prop}
		\label{prop:There-exist-positive-1}There exists a positive constant
		$\mathcal{T}^{*}$ and small enough $\lambda_{j}(j=2,...,5)$ such
		that for $\forall0<\mathcal{T}_{i}\leq\mathcal{T}{}^{*}$
		\begin{equation}
		\dot{\mathcal{V}}_{i}\leq-\overline{a}\mathcal{V}_{i}+\overline{\varepsilon}_{i}(t),\label{eq:32-1-1-1}
		\end{equation}
		where $i=1,2,...,N$, $\overline{a}>0$ is a  constant, $\overline{\varepsilon}_{i}(t)\in\mathrm{\mathbb{E}}(0)$.
	\end{prop}
	\begin{proof}
		See Appendix E.
	\end{proof}
	
	By solving the above inequality (\ref{eq:32-1-1-1}) for $V_{i}$,
	we can complete the proof of Theorem \ref{thm:1-1-1-1-1}.
	\begin{rem}
	\textcolor{black}{	The proposed results have several promising advantages compared with the existing works \cite{key-9-1,key-10-1,key-13-1,key-14}: 1) The proposed control scheme uses the PET (periodic event-triggered) feedback data through the network communication. This feature can automatically guarantee a Zeno-free behavior and avoid the continuous monitoring of triggered conditions. Moreover, in most real engineering situations, data transmission is often conducted periodically through the network. Therefore, PET mechanism is more suitable for real practical network control systems compared with continuous-time event triggered condition; 2) Our method deals with the COR (cooperative output regulation) problem for nonlinear MASs. This is more general than the traditional consensus problem for linear MASs \cite{key-13-1,key-14}. The proposed method aims to regulate the output of each follower in MASs to a given reference signal/leader while simultaneously rejecting the external disturbance; 3) Predictor feedback techniques are used to compensate for the input delay. This indicates that our method can handle long input delay in contrast with those passive methods \cite{key-9-1,key-10-1}.}
		
	\textcolor{black}{	The drawback of the proposed method may lie on that it needs extra computation for the predictor feedback terms in (21) compared with those passive methods \cite{key-9-1,key-10-1}. However, this issue will be alleviated as the development of high performance computers and chips. }
	\end{rem}

    \begin{rem}
\textcolor{black}{We will present some guidelines to select the control parameters. For the observer in Section III, the observer gains $\kappa_{1},\kappa_{2}$  and the sampling period $T^{ij}$ should be small enough. They should satisfy $\ensuremath{\kappa T\leq M}$ in Theorem 1. For the PET mechanism, large $\delta_{S}$,$\delta_{v}$ and small $\gamma_{S},\gamma_{v}$ can reduce the transmission times, while small $\delta_{S},\delta_{v}$ and large $\gamma_{S},\gamma_{v}$ will increase the communication burden. According to the above analysis, we can first select small $\kappa_{1},\kappa_{2},T^{ij}, \delta_{S},\delta_{v}$ and large $\gamma_{S},\gamma_{v}$  to guarantee the convergence of the observer. In this case, the behavior of the observer is like a continuous-time observer.  Next, one can increase $T^{ij}, \delta_{S},\delta_{v}$, and decrease $\gamma_{S},\gamma_{v}, \kappa_{1},\kappa_{2}$ to reduce the communication burden.}

\textcolor{black}{For the controller in Section IV, the control gain should satisfy $\ell+K_{i}<0$. For the PET controller in Section V, the sampling period $\mathcal{T}_{i}$ should be small enough and the control gain $|K_{i}|$ should be large enough. The selection of parameters $\mathcal{T}_{i}, K_{i}$ is similar to the observer in Section III. For the PET mechanism, large $\delta_{\phi}$ and small $\gamma_{\phi}$ can reduce the transmission times, while small $\delta_{\phi}$ and large $\gamma_{\phi}$ will increase the communication burden. }
    \end{rem}
	
	\section{Simulations}
	
	Consider nonlinear MASs (\ref{eq:1})-(\ref{eq:8-4}) with 1 leader
	and 4 followers. The communication graph is shown in Fig. \ref{fig:1-4}.
	For the leader, set $S=\mathrm{col}([0\thinspace1],[-1\thinspace0])$.
	Let $\dot{X}_{i}=f_{i}(X_{i})+U_{i}(t-D_{i})$
	where $f_{i}(X_{i})=X_{i}+0.1\mathrm{sin}(X_{i})$, $D_{i}=0.15s(i=1,2,3,4)$.

	
	
\begin{figure}
		\begin{centering}
			\includegraphics[scale=0.4]{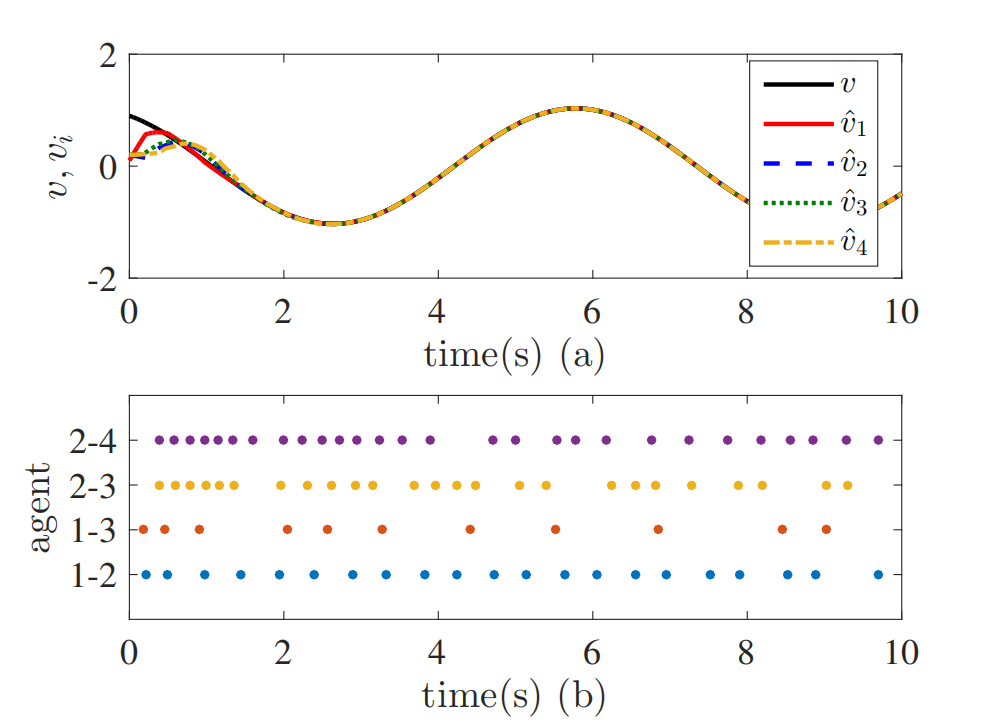}
			\par\end{centering}
		\caption{\label{fig:1}Performance of the proposed periodic event-triggered
			distributed observer. (a) Variations of the leader state $v$ and
			the estimation $\hat{v}_{i}(i=1,2,3,4)$; (b) periodic event-triggered
			instants for every agent pair.}
	\end{figure}

	The  distributed observer (\ref{eq:5-1-1-1})-(\ref{eq:6-1-1})
	and the controller (\ref{eq:45-1}) are utilized. The  parameters
	are set as: $\kappa_{1}=\kappa_{2}=3$, $K_{i}=-5(i=1,2,3,4)$. The
	sampling periods are set as $T^{01}=0.01s,T^{11}=0.01s,T^{12}=0.02s$;
	$T^{22}=0.01s,T^{24}=0.02s$; $T^{33}=0.02s,T^{34}=0.01s$. Fig. \ref{fig:1}(a)
	shows the performance of the periodic event-triggered adaptive distributed
	observer. We can see that the estimation values $\hat{v}_{i}$ of
	the 4 followers quickly track the true value $v$. Fig. \ref{fig:1}(b)
	shows the periodic event-triggered time instants between every two
	agents. We can see that there are a few data transmission times. It is less than one tenth of the transmission times with 10ms sampling period. This
	implies that the proposed periodic event-triggered condition can reduce
	the transmission times considerably. Moreover, the event-triggered
	time instants and sampling periods for each agent pair are not the
	same. This indicates that the data are transmitted in a fully asynchronous
	way.
	
\begin{figure}
		\begin{centering}
			\includegraphics[scale=0.4]{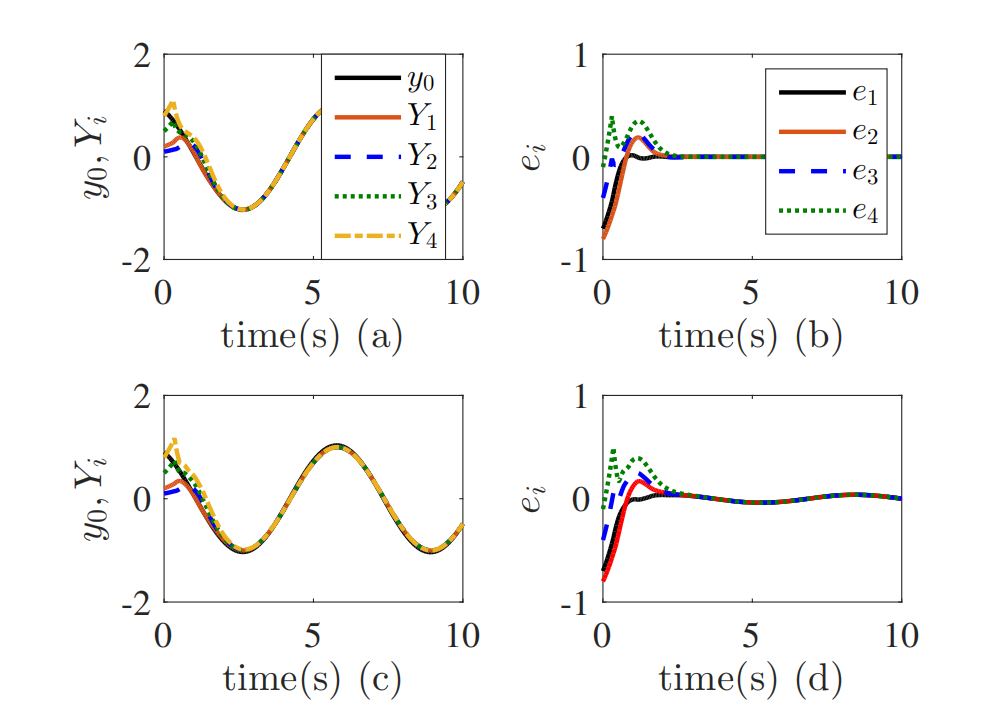}
			\par\end{centering}
		\caption{\label{fig:1-1}Control performance of the proposed controller when
			the switch is on node 1 in Fig. \ref{fig:1-3}. (a)-(b) Outputs and
			regulation error of the MASs when there is no delay mismatch; (c)-(d)
			Outputs and regulation error of the MASs when delay mismatch exists.}
	\end{figure}

\begin{figure}
		\begin{centering}
			\includegraphics[scale=0.4]{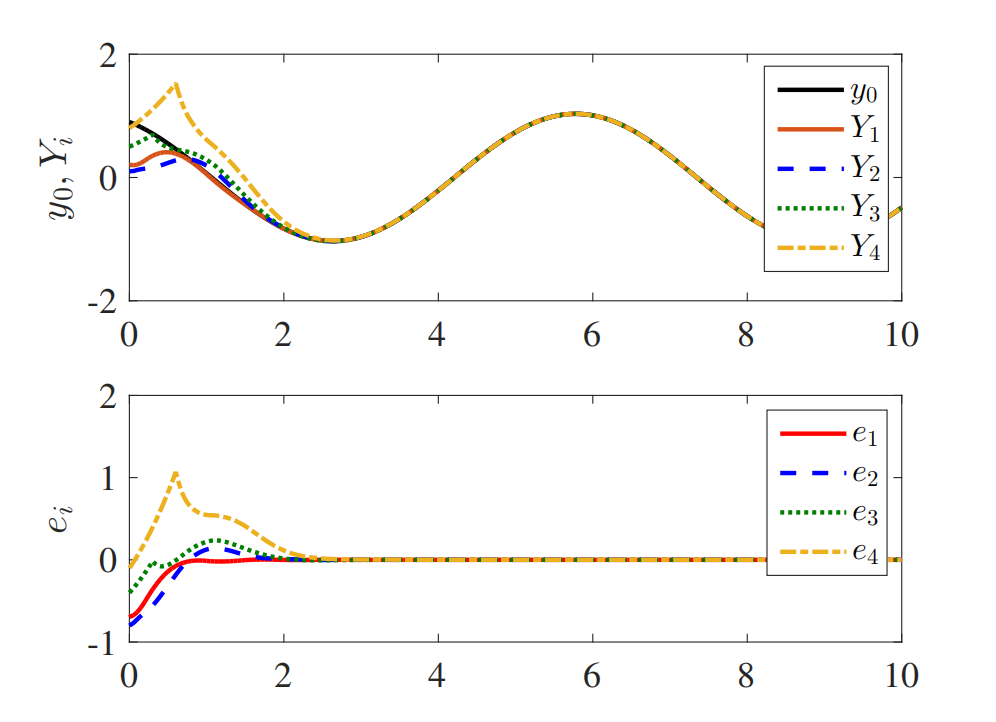}
			\par\end{centering}
		\caption{\label{fig:1-2-1}Control performance of the proposed controller with
			different input delays.}
	\end{figure}

	Fig. \ref{fig:1-1}(a) demonstrates the control performance of the
	proposed control scheme when the estimated delay is selected as $\hat{D}_{i}=D_{i}=0.15s$.
	We can see that the outputs of the followers quickly track the output
	of the leader. This demonstrates that $e_{i}(i=1,2,3,4)$ all converge
	to zero. If we take $\hat{D}_{i}=0.2s\neq D_{i}=0.15s$ implying that
	a delay mismatch exists, the tracking performance of the MASs is illustrated
	in Fig. \ref{fig:1-1}(b). We can see that $e_{i}(i=1,2,3,4)$ converge
	to a small set. Hence, the practical COR is achieved. 
	\textcolor{black}{Next, let the input delays for each agent be $D_{1}=0.01s, D_{2}=0.03s, D_{3}=0.15s, D_{4}=0.30s$ respectively. Fig. \ref{fig:1-2-1} shows the control performance when all the agents have different input delays. We can see that the actual outputs of the agents can track the leader very well. This indicates that the proposed results are effective.}

\begin{figure}
		\begin{centering}
			\includegraphics[scale=0.4]{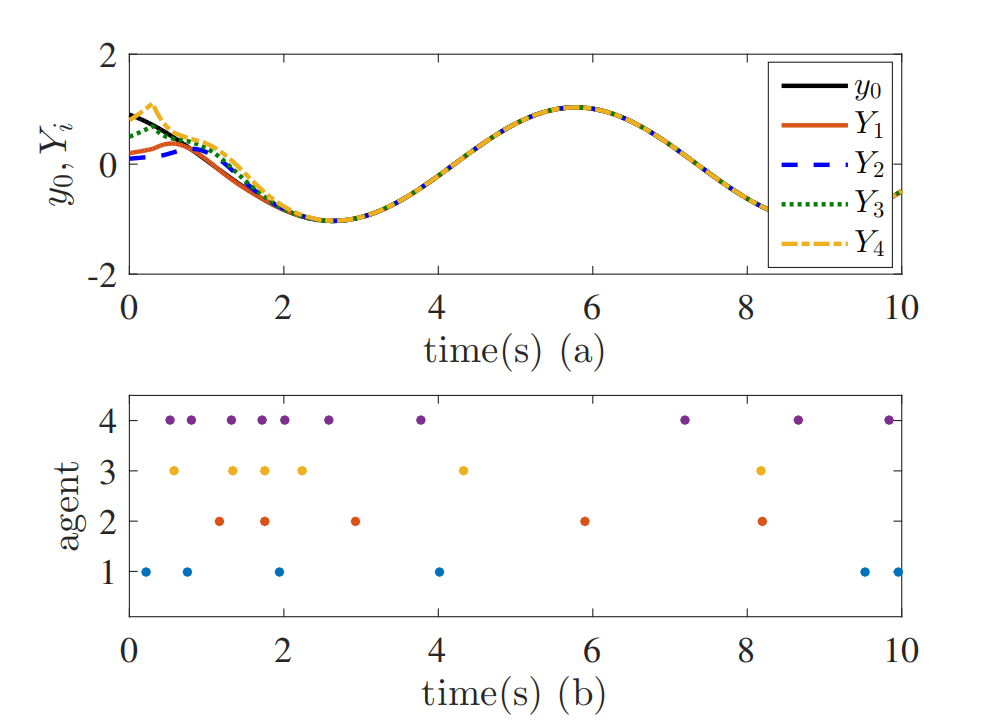}
			\par\end{centering}
		\caption{\label{fig:1-2}Control performance of the proposed controller when
			the switch is on node 2 in Fig. \ref{fig:1-3}. (a) Outputs of the
			MASs; (b) periodic event-triggered time instants between the sensor
			and the controller.}
	\end{figure}

 \textcolor{black}{	When PETM-B is invoked, the periodic event-triggered filter (\ref{eq:45-2})-(\ref{eq:23-2})
	is adopted with $L_{i}=-5(i=1,2,3,4)$. The controller is also in
	the form of (\ref{eq:45-1}). The  performance is demonstrated in
	Fig. \ref{fig:1-2}(a). We can see that the output regulation errors
	also converge to zero exponentially. Fig. \ref{fig:1-2}(b) illustrates
	the event-triggered time instants between the sensor and the controller.
	We can see that the communication resource has been saved.}
	
\begin{figure}
		\begin{centering}
			\includegraphics[scale=0.28]{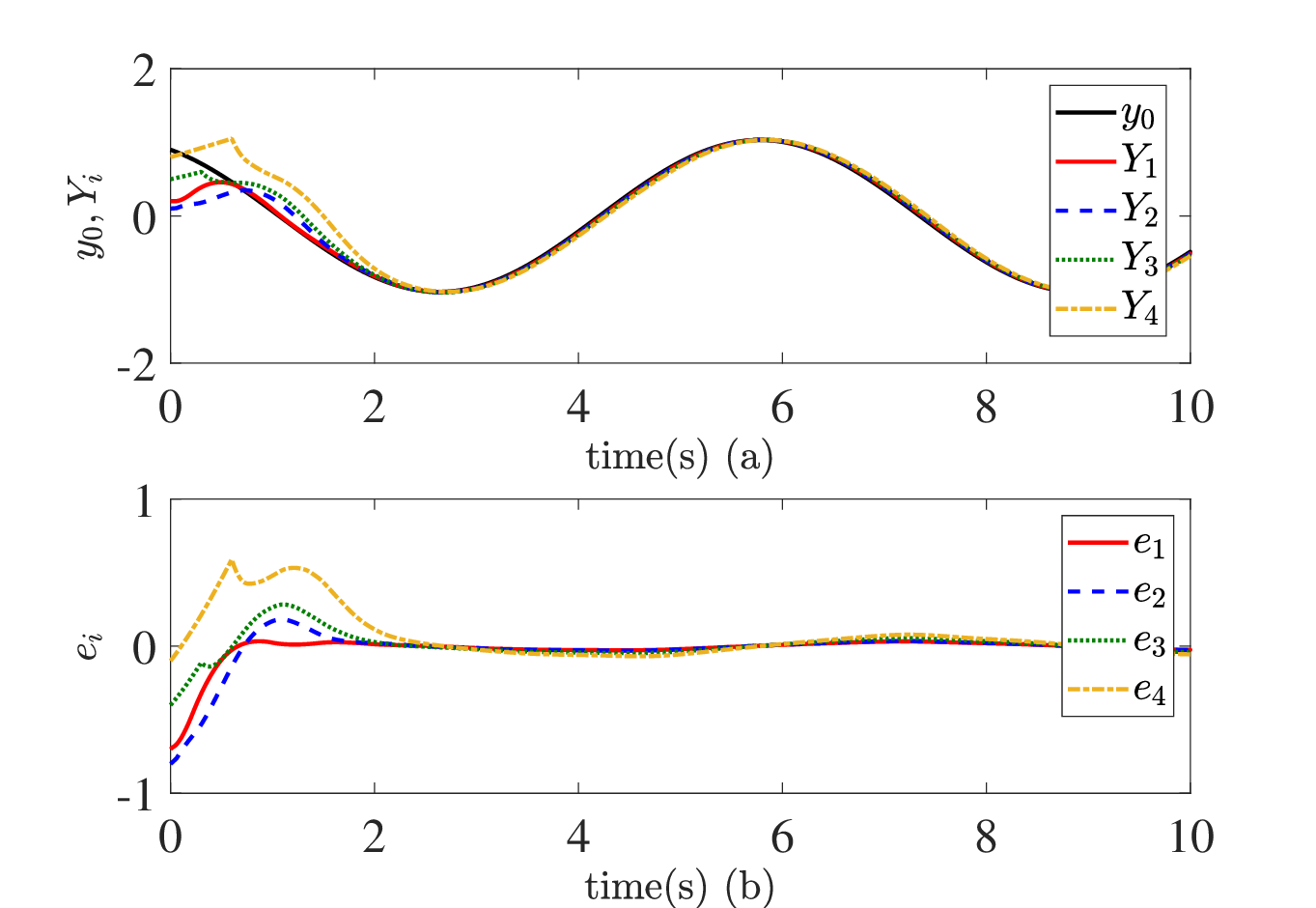}
			\par\end{centering}
		\caption{\label{fig:1-2-1-1}\textcolor{black}{Control performance of the proposed controller
			for complex nonlinear system with external disturbance.}}
	\end{figure}

\textcolor{black}{
Finally, we revise the models of the followers  into:
\begin{align*}
\dot{X}_{i} & =f_{i}(X_{i})+\Delta d_{i}+U_{i}(t-D_{i}),
\end{align*}
where $f_{i}(X_{i})=0.1\mathrm{sin}(X_{i})X_{i}+\cos^{2}(X_{i})\mathrm{ln}(1+X_{i}^{2})$,
$\Delta d_{i}=0.1\mathrm{sin}(t)+0.1\cos(t)$ is an unknown external
disturbance. The control performance is
shown in Fig. \ref{fig:1-2-1-1}. It can be seen that the output regulation errors
of the four followers can all converge to the neighborhood of the origin asymptotically. This
indicates that the proposed method can handle more complex nonlinear
systems, and is robust to external disturbance. }
	
	
	\section{Conclusions}
\textcolor{black}{	In this paper, we consider the periodic event-triggered COR problem
	for nonlinear MASs with long input delay. Based on fully asynchronous sampled data, a new
	predictor feedback controller is proposed to handle the long input
	delay in the nonlinear MASs. According to the simulation results, the transmission times of the proposed method is less than $10\%$ of the sampled data control with 10ms sampling period. Meanwhile, the sampling periods, input delays, and nonlinear models of each agent can be different from each other. In addition, the input delay can be arbitrary long in theory.}

	\section{Appendix}

\appendices{}

\subsection{Proof of Theorem 1}
\begin{proof}
	We will show for $\forall i=1,2,...,N$, $||\tilde{v}_{i}||\in\mathbb{E}(0)$
	if $||\tilde{S}_{i}||\in\mathbb{E}(0)$. The convergence of $\tilde{S}_{i}$
	can be proved similarly.
	
	Let
	\begin{align}
	z_{i}(t) & =\mathrm{e}^{-St}\hat{v}_{i}(t)(i=0,1,...,n),\label{eq:9-2-1}
	\end{align}
	where $z_{0}(t)=\mathrm{e}^{-St}\hat{v}_{0}(t)=\mathrm{e}^{-St}v(t)=v(0).$
	Then, (7) is written as:
	\begin{align}
	\dot{z}_{i}= & -\mathrm{e}^{-St}S\hat{v}_{i}(t)+\mathrm{e}^{-St}\hat{S}_{i}\hat{v}_{i}(t)\nonumber \\
	& +\mathrm{e}^{-St}\kappa_{2}\sum_{j=0}^{N}a_{ij}(\overline{v}_{j}(t,\overline{t}_{l}^{ij})-\overline{v}_{i}(t,\overline{t}_{l}^{ii})).\label{eq:40-3}
	\end{align}
	Next, we define a sampling time
	sequence $\{t_{0},t_{1},...,t_{k},...\}$ by collecting all the asynchronous
	sampling time instance $\{t_{k}^{ij}\}$. Based on Lemma 1-4),
	using the periodic event-triggered condition (9) and the
	fact that $\tilde{S}_{i}\triangleq\hat{S}_{i}-S$ converges to zero
	exponentially, (\ref{eq:40-3}) can be put in the following form
	\begin{align}
	\dot{\tilde{z}}= & -\sum_{i=1}^{N}\kappa_{2}A_{i1}\tilde{\boldsymbol{z}}_{i}(t_{k})+\sum_{i=1}^{N}\kappa_{2}A_{i2}\tilde{\boldsymbol{z}}_{i}(t_{k})\nonumber \\
	& +A_{3}\tilde{z}+\kappa_{2}A_{4}+A_{5},\label{eq:40-2}
	\end{align}
	where $\bar{z}=\mathrm{col}(z_{0},z_{0},...,z_{0})$, $z=\mathrm{col}(z_{1},z_{2},...,z_{N})$,
	$\tilde{z}=\mathrm{col}(\tilde{z}_{1},\tilde{z}_{2},...,\tilde{z}_{N})$
	with $\tilde{z}_{i}=z_{i}-z_{0}$, $\tilde{\boldsymbol{z}}_{i}(t_{k})=\mathrm{col}(\tilde{z}_{1}(t_{k}^{i1}),\tilde{z}_{2}(t_{k}^{i2}),...,\tilde{z}_{N}(t_{k}^{iN}))$,
	$A_{i1}=\mathcal{H}_{i}\varotimes I$. By a tedious but straightforward
	calculation, we can show that every element in matrices $A_{i2},A_{3},A_{4},A_{5}$
	is related with the estimation error $\tilde{S}_{i}$ and the PETM
	(9). Therefore, $||A_{i2}||,||A_{3}||,||A_{4}||,||A_{5}||\in\mathrm{\mathbb{E}}(0)(i=1,2,...,N).$
	$\mathcal{H}_{i}$ is constructed according to $\mathcal{H}$. It
	has the same dimension as $\mathcal{H}$. Meanwhile, its $i$th row
	is the same as $\mathcal{H}$ and other elements are set to zero.
	
	Let $d_{ij}(t)=t-t_{k}^{ij}(i,j=1,2,...,N)$ in (\ref{eq:40-2}),
	then (\ref{eq:40-2}) can be transformed into the following system
	with multiple time delays.
	\begin{align}
	\dot{\tilde{z}}= & -\sum_{i=1}^{N}\kappa_{2}A_{i1}\tilde{\boldsymbol{z}}_{id}+\sum_{i=1}^{N}\kappa_{2}A_{i2}\tilde{\boldsymbol{z}}_{id}\nonumber \\
	& +A_{3}\tilde{z}+\kappa_{2}A_{4}+A_{5},\label{eq:39-3-1}
	\end{align}
	where $\tilde{\boldsymbol{z}}_{id}=(\tilde{z}_{1}(t-d_{i1}(t)),\tilde{z}_{2}(t-d_{i2}(t)),...,\tilde{z}_{N}(t-d_{iN}(t)))^{\mathrm{T}}$.
	(\ref{eq:39-3-1}) can be written as:
	\begin{align}
	\dot{\tilde{z}}= & -\kappa_{2}A_{1}\tilde{z}+\kappa_{2}A_{2}\tilde{z}\nonumber \\
	& +\sum_{i=1}^{N}\kappa_{2}A_{i1}\eta_{i}-\sum_{i=1}^{N}\kappa_{2}A_{i2}\eta_{i}\nonumber \\
	& +A_{3}\tilde{z}+\kappa_{2}A_{4}+A_{5},\label{eq:27-4}
	\end{align}
	where $A_{1}=\sum_{i=1}^{N}A_{i1}=\mathcal{H}\varotimes I$, $A_{2}=\sum_{i=1}^{N}A_{i2}$,
	$\eta_{i}(t)=\mathrm{col}(\eta_{i1}(t),\eta_{i2}(t),...,\eta_{iN}(t))$
	with $\eta_{ij}(t)=\int_{t-d_{ij}(t)}^{t}\dot{\tilde{z}}_{j}(s)ds$.
	
	Consider the following Lyapunov-Krasovskii function
	\begin{equation}
	V=\frac{1}{2}\tilde{z}^{\mathrm{T}}Q\tilde{z}+\int_{t-T}^{t}\int_{s}^{t}||\dot{\tilde{z}}(\theta)||^{2}d\theta ds,\label{eq:13-1}
	\end{equation}
	where $Q$ is a positive definite matrix satisfying $QA_{1}+A_{1}^{\mathrm{T}}Q=2I$.
	$Q$ exists due to the graph of the MASs contains a directed spanning
	tree. Note that (\ref{eq:13-1}) is an integration of
	the value $\dot{\tilde{z}}$. Though $\dot{\tilde{z}}$ exhibits jumps
	at corresponding event-triggered times, the Lyapunov function in (\ref{eq:13-1})
	is continuous.
	
	Using (\ref{eq:27-4}), $\dot{V}$ is computed as:
	\begin{align*}
	\dot{V}\leq & \tilde{z}^{\mathrm{T}}Q(-\kappa_{2}A_{1}\tilde{z}+\kappa_{2}A_{2}\tilde{z})\\
	& +\tilde{z}^{\mathrm{T}}Q\left(\sum_{i=1}^{N}\kappa_{2}A_{i1}\eta_{i}-\sum_{i=1}^{N}\kappa_{2}A_{i2}\eta_{i}\right)\\
	& +\tilde{z}^{\mathrm{T}}Q(A_{3}\tilde{z}+\kappa_{2}A_{4}+A_{5})\\
	& -\int_{t-T}^{t}||\dot{\tilde{z}}(s)||^{2}ds+T||\dot{\tilde{z}}(t)||^{2}.
	\end{align*}
	It follows that 
	\begin{align*}
	\dot{V}\leq & -\kappa_{2}||\tilde{z}||^{2}+\tilde{z}^{\mathrm{T}}(\kappa_{2}QA_{2}+QA_{3})\tilde{z}\\
	& +\tilde{z}^{\mathrm{T}}Q\left(\sum_{i=1}^{N}\kappa_{2}A_{i1}\eta_{i}-\sum_{i=1}^{N}\kappa_{2}A_{i2}\eta_{i}\right)\\
	& +\tilde{z}^{\mathrm{T}}Q(\kappa_{2}A_{4}+A_{5})\\
	& -\int_{t-T}^{t}||\dot{\tilde{z}}(s)||^{2}ds+T||\dot{\tilde{z}}(t)||^{2}.
	\end{align*}
	Noting that $||A_{i2}||,||A_{3}||,||A_{4}||,||A_{5}||\in\mathrm{\mathbb{E}}(0)(i=1,2,...,N)$
	and using Lemma 1-1) and 2), we have
	\begin{align}
	\dot{V}\leq & -\kappa_{2}||\tilde{z}||^{2}+\kappa_{2}c_{1}\mathrm{e}^{-\gamma t}||\tilde{z}||^{2}\nonumber \\
	& +\frac{\kappa_{2}}{4}||\tilde{z}||^{2}+\kappa_{2}N\sum_{i=1}^{N}||QA_{i1}||^{2}\cdot||\eta_{i}||^{2}\nonumber \\
	& +\frac{\kappa_{2}}{4}||\tilde{z}||^{2}+\kappa_{2}c_{1}\mathrm{e}^{-\gamma t}\sum_{i=1}^{N}||\eta_{i}||^{2}\nonumber \\
	& +\frac{\kappa_{2}}{4}||\tilde{z}||^{2}+\kappa_{2}c_{1}\mathrm{e}^{-\gamma t}\nonumber \\
	& +T||\dot{\tilde{z}}(t)||^{2}-\int_{t-T}^{t}||\dot{\tilde{z}}(s)||^{2}ds,\label{eq:29-4}
	\end{align}
	where $c_{1},\gamma$ are positive constants. Meanwhile, using (\ref{eq:27-4})
	for $||\dot{\tilde{z}}(t)||^{2}$ and Lemma 1-2)
	\begin{align}
	||\dot{\tilde{z}}(t)||^{2}\leq & (2N+5)\left(\kappa_{2}^{2}||A_{1}||^{2}||\tilde{z}||^{2}+c_{2}\kappa_{2}^{2}\mathrm{e}^{-\gamma t}||\tilde{z}||^{2}\right)\nonumber \\
	& +(2N+5)\sum_{i=1}^{N}\kappa_{2}^{2}||A_{i1}||^{2}||\eta_{i}||^{2}\nonumber \\
	& +(2N+5)\sum_{i=1}^{N}c_{2}\kappa_{2}^{2}\mathrm{e}^{-\gamma t}||\eta_{i}||^{2}\nonumber \\
	& +(2N+5)c_{2}\kappa_{2}^{2}\mathrm{e}^{-\gamma t},\label{eq:46}
	\end{align}
	where $c_{2}$ is a positive constant. For $\eta_{i}$, by Cauchy-Schwarz
	inequality, we have
	\[
	\eta_{ij}^{2}(t)=\left(\int_{t-d_{ij}(t)}^{t}\dot{\tilde{z}}_{j}(s)ds\right)^{2}\leq T\int_{t-T}^{t}||\dot{\tilde{z}}_{j}(s)||^{2}ds,
	\]
	where $j=1,2,...,n$, then
	\begin{equation}
	||\eta_{i}||^{2}\leq T\int_{t-T}^{t}||\dot{\tilde{z}}(s)||^{2}ds.\label{eq:12-2}
	\end{equation}
	Substituting (\ref{eq:46}) and (\ref{eq:12-2}) into (\ref{eq:29-4}),
	we get
	\begin{align*}
	\dot{V}\leq & -(\varGamma_{1}+\varGamma_{2})+\epsilon,
	\end{align*}
	where $\epsilon\in\mathrm{\mathbb{E}}(0)$,
	\begin{align*}
	\varGamma_{1}= & \left(\frac{\kappa_{2}}{8}-T(2N+5)\kappa_{2}^{2}||A_{1}||^{2}\right)||\tilde{z}||^{2}\\
	& +\left(\frac{\kappa_{2}}{8}-\kappa_{2}c_{1}\mathrm{e}^{-\gamma t}-T(2N+5)c_{2}\kappa_{2}^{2}\mathrm{e}^{-\gamma t}\right)||\tilde{z}||^{2},
	\end{align*}
	\begin{align*}
	\varGamma_{2}= & \left(\frac{1}{3}-T\kappa_{2}N\sum_{i=1}^{N}||QA_{i1}||^{2}\right)\int_{t-T}^{t}||\dot{\tilde{z}}(s)||^{2}ds\\
	& +\left(\frac{1}{3}-T^{2}(2N+5)\kappa_{2}^{2}\sum_{i=1}^{N}||A_{i1}||^{2}\right)\int_{t-T}^{t}||\dot{\tilde{z}}(s)||^{2}ds\\
	& +\left(\frac{1}{3}-NT\kappa_{2}c_{1}\mathrm{e}^{-\gamma t}-T^{2}(2N+5)N\kappa_{2}c_{2}\mathrm{e}^{-\gamma t}\right)\\
	& \cdot\int_{t-T}^{t}||\dot{\tilde{z}}(s)||^{2}ds,
	\end{align*}
	\[
	\epsilon=\kappa_{2}c_{1}\mathrm{e}^{-\gamma t}+T(2N+5)c_{2}\kappa_{2}^{2}\mathrm{e}^{-\gamma t}.
	\]
	Note that for any finite $T$, there exists a finite time instant
	$t_{0}$ such that
	\[
	\frac{\kappa_{2}}{8}-\kappa_{2}c_{1}\mathrm{e}^{-\gamma t_{0}}-T(2N+5)c_{2}\kappa_{2}^{2}\mathrm{e}^{-\gamma t_{0}}>0,
	\]
	\[
	\frac{1}{3}-NT\kappa_{2}c_{1}\mathrm{e}^{-\gamma t}-T{}^{2}(2N+5)N\kappa_{2}c_{2}\mathrm{e}^{-\gamma t}>0.
	\]
	Therefore, if $T$ satisfies
	\begin{equation}
	0<T<M/\kappa_{2},
	\end{equation}
	where $M=\min\{M_{1},M_{2},M_{3}\}$ with 
	\[
	M_{1}=\frac{1}{8(2N+5)||A_{1}||^{2}},
	\]
	\[
	M_{2}=\frac{1}{3N\sum_{i=1}^{N}||QA_{i1}||^{2}},
	\]
	\[
	M_{3}=\sqrt{\frac{1}{3(2N+5)\sum_{i=1}^{N}||A_{i1}||^{2}}}.
	\]
	Then 
	\begin{align}
	\dot{V}\leq & -c_{3}V+\epsilon\label{eq:43-1}
	\end{align}
	for $\forall t\in[t_{0},+\infty)$ where $c_{3}$ is a positive constant.
	Since (\ref{eq:40-2}) is a linear system, $V$ does not exhibit finite-time
	escape on $[0,t_{0})$. Therefore, $V(t_{0})$ is bounded. Then by
	solving the above inequality, we can show $||\tilde{v}_{i}||=||\mathrm{e}^{-St}\tilde{v}_{i}||\leq||\tilde{z}||\in\mathrm{\mathbb{E}}(0)$. 
\end{proof}

\subsection{Proof of Proposition 2}

The proof is divided into three parts:

\textit{1) }$\overline{X}$-\textit{dynamics:}

From (12) and (22), we know
\begin{align*}
\dot{\overline{X}}_{i}= & \overline{f}_{i}(\overline{X}_{i},v)+\overline{u}^{i}(0,t)\\
= & \overline{f}_{i}(\overline{X}_{i},v)+\hat{u}^{i}(0,t)+\widetilde{u}^{i}(0,t)\\
= & \overline{f}_{i}(\overline{X}_{i},v)+K_{i}\hat{\chi}^{i}(0,t)+\widetilde{u}^{i}(0,t)+\hat{w}^{i}(0,t).
\end{align*}
Due to $\hat{\chi}^{i}(0,t)=\hat{X}_{i}(t)$ by (20), it
follows that
\begin{align*}
\dot{\overline{X}}_{i}= & \overline{f}_{i}(\overline{X}_{i},v)+K_{i}\overline{X}_{i}(t)+\widetilde{u}^{i}(0,t)+\hat{w}^{i}(0,t)+\zeta_{i0}(t),
\end{align*}
where $\zeta_{i0}(t)=K_{i}\hat{X}_{i}(t)-K_{i}\overline{X}_{i}(t)=K_{i}F_{i}v-K_{i}F_{i}\hat{v}_{i}$.
By Theorem 1, we know $\zeta_{i0}(t)\in\mathrm{\mathbb{E}}(0)$.
(23) is proved.

\textit{2) }$u$-\textit{dynamics:}

Recall the definitions of $u^{i}(x,t),\overline{u}^{i}(x,t),\check{u}^{i}(x,t),\hat{u}^{i}(x,t)$
in (14), (15), (17) and (18).
Using the leader dynamics (1), we can obtain
\begin{equation}
D_{i}\overline{u}_{t}^{i}(x,t)=\overline{u}_{x}^{i}(x,t),\label{eq:36}
\end{equation}
\begin{equation}
\hat{D}_{i}\check{u}_{t}^{i}(x,t)=\check{u}_{x}^{i}(x,t),\label{eq:37}
\end{equation}
\begin{align}
& \hat{D}_{i}\hat{u}_{t}^{i}(x,t)=\hat{D}_{i}\check{u}_{t}^{i}(x,t)-\hat{D}_{i}\frac{d(R_{i}(\mathrm{e}^{\hat{S}_{i}\hat{D}_{i}x}\hat{v}_{i},\hat{S}_{i}))}{dt}\label{eq:33}
\end{align}
\begin{equation}
\hat{u}_{x}^{i}(x,t)=\check{u}_{x}^{i}(x,t)-\frac{d(R_{i}(\mathrm{e}^{\hat{S}_{i}\hat{D}_{i}x}\hat{v}_{i},\hat{S}_{i}))}{dx}.\label{eq:39}
\end{equation}
(\ref{eq:37})-(\ref{eq:39}) indicate that 
\begin{equation}
\hat{D}_{i}\hat{u}_{t}^{i}(x,t)=\hat{u}_{x}^{i}(x,t)+\Delta_{i1}(x,t),\label{eq:40-1}
\end{equation}
\begin{equation}
D_{i}\hat{u}_{t}^{i}(x,t)=\frac{D_{i}}{\hat{D}_{i}}\hat{u}_{x}^{i}(x,t)+\frac{D_{i}}{\hat{D}_{i}}\Delta_{i1}(x,t),\label{eq:40}
\end{equation}
where
\begin{align*}
& \Delta_{i1}(x,t)=\frac{d(R_{i}(\mathrm{e}^{\hat{S}_{i}\hat{D}_{i}x}\hat{v}_{i},\hat{S}_{i}))}{dx}-\hat{D}_{i}\frac{d(R_{i}(\mathrm{e}^{\hat{S}_{i}\hat{D}_{i}x}\hat{v}_{i},\hat{S}_{i}))}{dt}.
\end{align*}
Due to $\widetilde{u}^{i}(x,t)=\overline{u}^{i}(x,t)-\hat{u}^{i}(x,t)$,
using (\ref{eq:36}) and (\ref{eq:40}) we have (24)-(25)
where $\zeta_{i1}(x,t)=-\frac{D_{i}}{\hat{D}_{i}}\Delta_{i1}(x,t)$. 

Next, we show $\zeta_{i1}(x,t)\in\mathrm{\mathbb{E}}(0)$. We only
need show $\Delta_{i1}(x,t)\in\mathrm{\mathbb{E}}(0)$. Let $m_{i}(x,t)=\mathrm{e}^{\hat{S}_{i}\hat{D}_{i}x}\hat{v}_{i}$,
then
\begin{align}
\frac{d(R_{i}(\mathrm{e}^{\hat{S}_{i}\hat{D}_{i}x}\hat{v}_{i},\hat{S}_{i}))}{dx} & =\frac{d(R_{i}(m_{i},\hat{S}_{i}))}{dx}\nonumber \\
& =\frac{\partial R_{i}(m_{i},\hat{S}_{i})}{\partial m_{i}}\hat{S}_{i}\hat{D}_{i}\mathrm{e}^{\hat{S}_{i}\hat{D}_{i}x}\hat{v}_{i},\label{eq:37-2}
\end{align}

\begin{align}
& \hat{D}_{i}\frac{d(R_{i}(\mathrm{e}^{\hat{S}_{i}\hat{D}_{i}x}\hat{v}_{i},\hat{S}_{i}))}{dt}\nonumber \\
= & \hat{D}_{i}\frac{\partial R_{i}(m^{i},\hat{S}_{i})}{\partial m_{i}}\frac{dm_{i}}{dt}+\hat{D}_{i}\frac{\partial R_{i}(m_{i},\hat{S}_{i})}{\partial\hat{S}_{i}}\frac{d\hat{S}_{i}}{dt}\nonumber \\
= & \hat{D}_{i}\frac{\partial R_{i}(m_{i},\hat{S}_{i})}{\partial m_{i}}\mathrm{e}^{\hat{S}_{i}\hat{D}_{i}x}\hat{S}_{i}\hat{v}_{i}+\Delta_{i2}(x,t),\label{eq:38}
\end{align}
where 
\begin{align*}
& \Delta_{i2}(x,t)\\
= & \hat{D}_{i}\frac{\partial R_{i}(m_{i},\hat{S}_{i})}{\partial m_{i}}\mathrm{e}^{\hat{S}_{i}\hat{D}_{i}x}(\kappa_{2}\sum_{j=0}^{N}a_{ij}(\overline{v}_{j}(t,\overline{t}_{l}^{ij})-\overline{v}_{i}(t,\overline{t}_{l}^{ii})))\\
& +\hat{D}_{i}^{2}\frac{\partial R_{i}(m_{i},\hat{S}_{i})}{\partial m_{i}}\mathrm{e}^{\hat{S}_{i}\hat{D}_{i}x}\hat{S}_{i}(\kappa_{1}\sum_{j=0}^{N}a_{ij}(\hat{S}_{j}(\overline{t}_{l}^{ij})-\hat{S}_{i}(\overline{t}_{l}^{ii})))\hat{v}_{i}\\
& +\hat{D}_{i}\frac{\partial R_{i}(m_{i},\hat{S}_{i})}{\partial\hat{S}_{i}}(\kappa_{1}\sum_{j=0}^{N}a_{ij}(\hat{S}_{j}(\overline{t}_{l}^{ij})-\hat{S}_{i}(\overline{t}_{l}^{ii}))).
\end{align*}
Then, subtracting (\ref{eq:38}) from (\ref{eq:37-2}), we have, 
\[
\Delta_{i1}(x,t)=-\Delta_{i2}(x,t).
\]
From Theorem 1, we can conclude that $\Delta_{i2}(x,t)\in\mathrm{\mathbb{E}}(0)$
for $\forall x\in[0,1]$. Hence, $\Delta_{i1}(x,t),\zeta_{i1}(x,t)\in\mathrm{\mathbb{E}}(0)$.
(24)-(25) are proved.

\textit{3) }$w$-\textit{dynamics and }$w_{x}$-\textit{dynamics:}

By the backstepping transformation (22), we have
\begin{align*}
\hat{D}_{i}\hat{w}_{t}^{i}(x,t)= & \hat{D}_{i}\hat{u}_{t}^{i}(x,t)-\hat{D}_{i}K_{i}\hat{\chi}_{t}^{i}(x,t),
\end{align*}
\[
\hat{w}_{x}^{i}(x,t)=\hat{u}_{x}^{i}(x,t)-K_{i}\hat{\chi}_{x}^{i}(x,t).
\]
By (\ref{eq:40-1}), it follows that
\begin{align}
& \hat{D}_{i}\hat{w}_{t}^{i}(x,t)-\hat{w}_{x}^{i}(x,t)\nonumber \\
= & -K_{i}(\hat{D}_{i}\hat{\chi}_{t}(x,t)-\hat{\chi}_{x}(x,t))+\Delta_{i1}(x,t).\label{eq:39-2}
\end{align}
Next, some analysis will be conducted for $\hat{D}_{i}\hat{\chi}_{t}^{i}(x,t)-\hat{\chi}_{x}^{i}(x,t)$.
According to (20), we have 
\begin{align*}
& \hat{\chi}_{t}^{i}(x,t)\\
= & (\overline{f}_{i}(\overline{X}_{i},v)+K_{i}\overline{X}_{i}+\widetilde{u}^{i}(0,t)+\hat{w}^{i}(0,t)+\overline{\zeta}_{i0}(t))\\
& +\hat{D}_{i}\int_{0}^{x}\frac{\partial\overline{f}_{i}}{\partial\hat{\chi}^{i}}(y,t)\hat{\chi}_{t}^{i}(y,t)+\frac{\partial\overline{f}_{i}}{\partial\overline{m}^{i}}(y,t)m_{t}^{i}(y,t)\\
& +\hat{D}_{i}\int_{0}^{x}\hat{u}_{t}^{i}(y,t)dy,
\end{align*}
where $\overline{m}^{i}(x,t)=\mathrm{e}^{\hat{S}_{i}\hat{D}_{i}y}\hat{S}_{i}\hat{v}_{i}$,
$\overline{\zeta}_{i0}(t)\in\mathrm{\mathbb{E}}(0)$.
\begin{align*}
& \hat{\chi}_{x}^{i}(x,t)\\
= & \hat{D}_{i}(\overline{f}_{i}(\hat{\chi}^{i}(x,t),\overline{m}^{i}(x,t))+\hat{u}^{i}(x,t))\\
= & \hat{D}_{i}(\overline{f}_{i}(\hat{\chi}^{i}(0,t),\hat{v}_{i}(t))+\hat{u}^{i}(0,t))\\
& +\hat{D}_{i}\int_{0}^{x}\frac{\partial\overline{f}_{i}}{\partial\hat{\chi}^{i}}(y,t)\hat{\chi}_{y}^{i}(y,t)+\frac{\partial\overline{f}_{i}}{\partial\overline{m}^{i}}(y,t)\overline{m}_{y}^{i}(y,t)dy\\
& +\hat{D}_{i}\int_{0}^{x}\hat{u}_{y}^{i}(y,t)dy.
\end{align*}
Then, we have
\begin{align*}
& \hat{D}_{i}\hat{\chi}_{t}^{i}(x,t)-\hat{\chi}_{x}^{i}(x,t)\\
= & \hat{D}_{i}\int_{0}^{x}\frac{\partial\overline{f}_{i}}{\partial\hat{\chi}^{i}}(y,t)\left(\hat{D}_{i}\hat{\chi}_{t}^{i}(y,t)-\hat{\chi}_{y}^{i}(y,t)\right)dy\\
& +\hat{D}_{i}\int_{0}^{x}\frac{\partial\overline{f}_{i}}{\partial\overline{m}^{i}}(y,t)\left(\hat{D}_{i}\overline{m}_{t}^{i}(y,t)-\overline{m}_{y}^{i}(y,t)\right)dy\\
& +\hat{D}_{i}\int_{0}^{x}\left(\hat{D}_{i}\hat{u}_{t}^{i}(y,t)-\hat{u}_{y}^{i}(y,t)\right)dy\\
& -\hat{D}_{i}(\overline{f}_{i}(\hat{X}_{i},\hat{v}_{i})+\hat{u}^{i}(0,t)).\\
& +\hat{D}_{i}(\overline{f}_{i}(\overline{X}_{i},v)+K_{i}\overline{X}_{i}+\widetilde{u}^{i}(0,t)+\hat{w}^{i}(0,t)+\overline{\zeta}_{i0}(t)).
\end{align*}
It follows that
\begin{align*}
& \hat{D}_{i}\hat{\chi}_{t}^{i}(x,t)-\hat{\chi}_{x}^{i}(x,t)\\
= & \hat{D}_{i}\int_{0}^{x}\frac{\partial\overline{f}_{i}}{\partial\hat{\chi}^{i}}(y,t)\left(\hat{D}_{i}\hat{\chi}_{t}^{i}(y,t)-\hat{\chi}_{x}^{i}(y,t)\right)dy\\
& +\hat{D}_{i}\widetilde{u}^{i}(0,t)+\Delta_{i3}(x,t),
\end{align*}
where $\Delta_{i3}(x,t)\in\mathrm{\mathbb{E}}(0)$.

Let $\overline{\chi}^{i}(x,t)=\hat{D}_{i}\hat{\chi}_{t}^{i}(x,t)-\hat{\chi}_{x}^{i}(x,t)$.
Then, we have
\begin{align*}
\overline{\chi}_{x}^{i}(x,t) & =\hat{D}_{i}\frac{\partial\overline{f}_{i}}{\partial\hat{\chi}^{i}}(x,t)\overline{\chi}^{i}(x,t)+\frac{\partial\Delta_{i3}(x,t)}{\partial x}\\
\overline{\chi}^{i}(0,t) & =\hat{D}_{i}\widetilde{u}^{i}(0,t)+\Delta_{i3}(0,t).
\end{align*}
By solving the above equation with respect to the space $x$, we get
\begin{align*}
\overline{\chi}^{i}(x,t)= & \hat{D}_{i}\widetilde{u}^{i}(0,t)\Theta^{i}(x,t)+\Delta_{i4}(x,t),
\end{align*}
where $\Delta_{i4}(x,t)\in\mathrm{\mathbb{E}}(0)$,
\[
\Theta^{i}(x,t)=\mathrm{e}^{\intop_{0}^{x}\hat{D}_{i}\frac{\partial\overline{f}_{i}}{\partial\hat{\chi}^{i}}(y,t)dy}
\]
Using this for (\ref{eq:39-2}), we can prove (26)-(27).

$w_{x}$-dynamics can be obtained by differentiating $w$ with respect
to $x$. 

\subsection{Proof of Proposition 3}

We will analyze each terms in the Lyapunov function (30). 

For $V_{i1}$, using (23) and Young's inequality, we can
obtain
\begin{align}
\dot{V}_{i1}\leq & \overline{X}_{i}(\overline{f}_{i}(\overline{X}_{i},v)+K_{i}\overline{X}_{i}+\widetilde{u}^{i}(0,t)+\hat{w}^{i}(0,t)+\zeta_{i0}(t))\nonumber \\
\leq & (K_{i}+\ell+1)||\overline{X}_{i}||^{2}\nonumber \\
& +c_{1}\widetilde{u}^{i}(0,t)^{2}+c_{1}w^{i}(0,t)^{2}+c_{1}\varepsilon_{i1}^{2},\label{eq:44-3}
\end{align}
where $c_{1}$ is a positive constant, $\varepsilon_{i1}(t)\in\mathrm{\mathbb{E}}(0)$.

For $V_{i2}$, using (24)-(25) and integration
by parts, the derivative of $V_{i2}$ is expressed as
\begin{align*}
\dot{V}_{i2}= & -2\widetilde{D}_{i}\int_{0}^{1}\widetilde{u}^{i}(x,t)\frac{\hat{u}_{x}^{i}(x,t)}{\hat{D}_{i}}(1+x)dx\\
& +2\int_{0}^{1}\widetilde{u}^{i}(x,t)\zeta_{i1}(x,t)(1+x)dx\\
& +2\widetilde{u}^{i}(1,t)^{2}-\widetilde{u}^{i}(0,t)^{2}-\int_{0}^{1}\widetilde{u}^{i}(x,t)^{2}dx.
\end{align*}
Note that from (22) and (20),
\begin{align}
\hat{u}_{x}^{i}(x,t)= & \hat{w}_{x}^{i}(x,t)+K_{i}\hat{\chi}_{x}^{i}(x,t),\nonumber \\
\hat{\chi}_{x}^{i}(x,t)= & \hat{D}_{i}\overline{f}_{i}(\hat{\chi}^{i}(x,t),\mathrm{e}^{\hat{S}_{i}\hat{D}_{i}x}\hat{S}_{i}\hat{v}_{i})\nonumber \\
& +\hat{D}_{i}\hat{w}^{i}(x,t)+\hat{D}_{i}\widetilde{u}^{i}(x,t)+\hat{D}_{i}K_{i}\hat{\chi}^{i}(x,t),\label{eq:59}
\end{align}
\begin{align}
& \hat{\chi}^{i}(x,t)\nonumber \\
= & \hat{X}_{i}(t)+\hat{D}_{i}\int_{0}^{x}\overline{f}_{i}(\hat{\chi}^{i}(y,t),\mathrm{e}^{\hat{S}_{i}\hat{D}_{i}y}\hat{S}_{i}\hat{v}_{i})dy\nonumber \\
& +\hat{D}_{i}\int_{0}^{x}(\hat{w}^{i}(y,t)+\widetilde{u}^{i}(y,t)+K_{i}\hat{\chi}^{i}(y,t))dy.\label{eq:60}
\end{align}
Then, noting that $|\overline{f}_{i}(\hat{\chi}^{i}(x,t),\mathrm{e}^{\hat{S}_{i}\hat{D}_{i}x}\hat{S}_{i}\hat{v}_{i})|\leq\ell|\hat{\chi}^{i}(x,t)|$
and based on the Gronwall inequality, (\ref{eq:60}) implies that
\begin{align*}
|\hat{\chi}^{i}(x,t)|\leq & \sigma_{1}||\hat{X}_{i}||++\sigma_{1}\int_{0}^{1}|\hat{w}^{i}(x,t)|dx\\
& +\sigma_{1}\int_{0}^{1}|\widetilde{u}^{i}(x,t)|dx,
\end{align*}
where $\sigma_{1}$ is a positive constant.

Similarly, by resorting to (\ref{eq:59}), we can show
\begin{align*}
|\hat{\chi}_{x}^{i}(x,t)|\leq & \sigma_{2}||\hat{X}_{i}||+\sigma_{2}\int_{0}^{1}|\widetilde{u}^{i}(x,t)|dx\\
& +\sigma_{2}\int_{0}^{1}|\hat{w}^{i}(x,t)|dx\\
& +\sigma_{2}|\hat{w}^{i}(x,t)|+\sigma_{2}|\widetilde{u}^{i}(x,t)|,
\end{align*}
where $\sigma_{2}$ is a positive constant.

Then, using Young and Cauchy-Schwarz inequalities, the derivative
of $V_{i2}$ can be computed as
\begin{align}
\dot{V}_{i2}\leq & 2\widetilde{u}^{i}(1,t)^{2}-\frac{1}{2}\int_{0}^{1}\widetilde{u}^{i}(x,t)^{2}dx-\widetilde{u}^{i}(0,t)^{2}\nonumber \\
& +c_{2}|\widetilde{D}_{i}|\left(||\overline{X}_{i}||^{2}+\int_{0}^{1}\widetilde{u}^{i}(x,t)^{2}dx\right)\nonumber \\
& +c_{2}|\widetilde{D}_{i}|\left(\int_{0}^{1}\hat{w}^{i}(x,t)^{2}dx+\int_{0}^{1}\hat{w}_{x}^{i}(x,t)^{2}dx\right)\nonumber \\
& +c_{2}\varepsilon_{i2}^{2},\label{eq:44-2}
\end{align}
where $c_{2}$ is a positive constant, $|\widetilde{u}^{i}(1,t)|\in\mathrm{\mathbb{E}}(\widetilde{D}_{i})$,
$\varepsilon_{i2}(t)\in\mathrm{\mathbb{E}}(0)$.

For $V_{i3}$, using (26)-(27) and integration
by parts, the derivative of $V_{i3}$ is written as
\begin{align}
\dot{V}_{i3}= & -\hat{w}^{i}(0,t)^{2}-\int_{0}^{1}\hat{w}^{i}(x,t)^{2}dx\nonumber \\
& +2\int_{0}^{1}\hat{w}^{i}(x,t)(\hat{D}_{i}\widetilde{u}^{i}(0,t)\Theta^{i}(x,t))(1+x)dx\nonumber \\
& +2\int_{0}^{1}(1+x)\zeta_{i3}(x,t)dx\nonumber \\
\leq & -\hat{w}^{i}(0,t)^{2}-\frac{1}{2}\int_{0}^{1}\hat{w}^{i}(x,t)^{2}dx\nonumber \\
& +c_{3}\widetilde{u}^{i}(0,t)^{2}+c_{3}\varepsilon_{i3}^{2},\label{eq:44-1}
\end{align}
where $c_{3}$ is a positive constant, $\varepsilon_{i3}(t)\in\mathrm{\mathbb{E}}(0)$.

Similarly for $V_{i4}$, we can obtain
\begin{align}
\dot{V}_{i4}= & 2\hat{w}_{x}^{i}(1,t)^{2}-\int_{0}^{1}\hat{w}_{x}^{i}(x,t)^{2}dx-\hat{w}_{x}^{i}(0,t)^{2}\nonumber \\
& +2\int_{0}^{1}\hat{w}_{x}^{i}(x,t)\hat{D}_{i}\widetilde{u}^{i}(0,t)\Theta_{x}^{i}(x,t)(1+x)dx\nonumber \\
& +2\int_{0}^{1}\hat{w}_{x}^{i}(x,t)\zeta_{i4}(x,t)(1+x)dx\nonumber \\
\leq & -\hat{w}_{x}^{i}(0,t)^{2}-\frac{1}{2}\int_{0}^{1}\hat{w}_{x}^{i}(x,t)^{2}dx\nonumber \\
& +c_{4}\widetilde{u}^{i}(0,t)^{2}+c_{4}\varepsilon_{i4}^{2},\label{eq:44}
\end{align}
where $c_{4}$ is a positive constant, $\varepsilon_{i4}(t)\in\mathrm{\mathbb{E}}(0)$.

According to (\ref{eq:44-3})-(\ref{eq:44}), the derivative of $V_{i}$
is given by:
\begin{align*}
\dot{V}_{i}\leq & \sum_{j=1}^{3}\Lambda_{ij}+\varepsilon_{i},
\end{align*}
where $\varepsilon_{i}(t)\in\mathrm{\mathbb{E}}(\widetilde{D}_{i})$,
\begin{align*}
\Lambda_{i1}=- & \left(K_{i}+\ell+1+\lambda_{2}|\widetilde{D}_{i}|c_{2}\right)||\overline{X}_{i}||^{2},
\end{align*}
\begin{align*}
\Lambda_{i2}= & -\left(\frac{\lambda_{2}}{2}-\lambda_{2}|\widetilde{D}_{i}|c_{2}\right)\int_{0}^{1}\widetilde{u}^{i}(x,t)^{2}dx\\
& -\left(\lambda_{2}-c_{1}\lambda_{1}-c_{3}\lambda_{3}-c_{4}\lambda_{4}\right)\widetilde{u}^{i}(0,t)^{2}\\
& -\left(\lambda_{3}-c_{1}\lambda_{1}\right)\hat{w}^{i}(0,t)^{2},
\end{align*}
\begin{align*}
\Lambda_{i3}= & -\left(\frac{\lambda_{3}}{2}-\lambda_{2}|\widetilde{D}_{i}|c_{2}\right)\int_{0}^{1}\hat{w}^{i}(x,t)^{2}dx\\
& -\left(\frac{\lambda_{4}}{2}-\lambda_{2}|\widetilde{D}_{i}|c_{2}\right)\int_{0}^{1}\hat{w}_{x}^{i}(x,t)^{2}dx.
\end{align*}
We can complete the proof.

\subsection{Proof of Proposition 4}

Subtracting (32) from (12), we obtain
\begin{align}
\dot{\widetilde{X}}_{i}= & \overline{f}_{i}(\overline{X}_{i},v)-\overline{f}_{i}(\hat{X}_{i},\hat{v}_{i})+L_{i}(\phi_{i}(\overline{\tau}_{q}^{i})-\hat{X}_{i}(\tau_{p}^{i})).\label{eq:33-1-1}
\end{align}
It is noted that for the term $\phi_{i}(\overline{\tau}_{q}^{i})-C_{i}\hat{X}_{i}(\overline{\tau}_{q}^{i})$,
using (26) we get
\begin{align*}
& \phi_{i}(\overline{\tau}_{q}^{i})-C_{i}\hat{X}_{i}(\tau_{p}^{i})\\
= & \phi_{i}(\tau_{p}^{i})-C_{i}\hat{X}_{i}(\tau_{p}^{i})+\phi_{i}(\overline{\tau}_{q}^{i})-\phi_{i}(\tau_{p}^{i})\\
= & Y_{i}(\tau_{p}^{i})-F_{i}v(\tau_{p}^{i})-C_{i}\hat{X}_{i}(\tau_{p}^{i})\\
& -F_{i}\overline{v}_{i}(\tau_{p}^{i},\overline{t}_{l}^{ii})+F_{i}v(\tau_{p}^{i})+\phi_{i}(\overline{\tau}_{q}^{i})-\phi_{i}(\tau_{p}^{i}).
\end{align*}
It follows that 
\begin{align}
\phi_{i}(\overline{\tau}_{q}^{i})-C_{i}\hat{X}_{i}(\tau_{p}^{i})=C_{i}\widetilde{X}_{i}(\tau_{p}^{i})+\overline{\Delta}_{i1}(t),\label{eq:29-2}
\end{align}
where $\overline{\Delta}_{i1}(t)=-F_{i}\overline{v}_{i}(\tau_{p}^{i},\overline{t}_{l}^{i})+F_{i}v(\tau_{p}^{i})+\phi_{i}(\overline{\tau}_{q}^{i})-\phi_{i}(\tau_{p}^{i}).$
According to the periodic event-triggered condition and Theorem 1,
we know $\overline{\Delta}_{i1}(t)\in\mathrm{\mathbb{E}}(0)$.

Using (\ref{eq:29-2}) for (\ref{eq:33-1-1}), we can obtain (36).
(37)-(39) can be proved similarly by resorting
to the proof Proposition 2.

\subsection{Proof of Proposition 5}

For $\mathcal{V}_{i1}$, by Lemma 1-1)
and (36), we have 
\begin{align*}
\dot{\mathcal{V}}_{i1}\leq & (L_{i}+\ell+1)||\widetilde{X}_{i}||^{2}\\
& +c_{1}||\widetilde{X}_{i}-\widetilde{X}_{i}(\tau_{p}^{i})||^{2}+c_{1}\overline{\varepsilon}_{i0}^{2},
\end{align*}
where $c_{1}$ is a positive constant, $\overline{\varepsilon}_{i0}\in\mathrm{\mathbb{E}}(0)$.

For $\mathcal{V}_{i2}$, by (37), we have 
\begin{align*}
\dot{\mathcal{V}}_{i2}\leq & (K_{i}+\ell+1)||\hat{X}_{i}||^{2}+c_{2}||\widetilde{X}_{i}||^{2}+c_{2}\hat{w}^{i}(0,t)^{2}\\
& +c_{2}||\widetilde{X}_{i}-\widetilde{X}_{i}(\tau_{p}^{i})||^{2}+c_{2}||\hat{X}_{i}-\hat{X}_{i}(\tau_{p}^{i})||^{2}\\
& +c_{2}\overline{\varepsilon}_{i0}^{2},
\end{align*}
where $c_{2}$ is a positive constant, $\overline{\varepsilon}_{i0}\in\mathrm{\mathbb{E}}(0)$.

For $\mathcal{V}_{i3}$, by integration by parts, we have
\begin{align*}
\dot{\mathcal{V}}_{i3}\leq & -\hat{w}^{i}(0,t)^{2}-\frac{1}{2}\int_{0}^{1}\hat{w}^{i}(x,t)^{2}dx\\
& +c_{3}||\widetilde{X}_{i}||^{2}+c_{3}||\widetilde{X}_{i}-\widetilde{X}(\tau_{p}^{i})||^{2}\\
& +c_{3}||\hat{X}_{i}-\hat{X}_{i}(\tau_{p}^{i})||^{2}\\
& +c_{3}\overline{\varepsilon}_{i3}^{2},
\end{align*}
where $c_{3}$ is a positive constant, $\overline{\varepsilon}_{i3}\in\mathrm{\mathbb{E}}(0)$.

For $\mathcal{V}_{i4}$, by (36) we have
\begin{align*}
\dot{\mathcal{V}}_{i4}\leq & c_{4}\mathcal{T}_{i}||\widetilde{X}_{i}||^{2}\\
& +c_{4}\mathcal{T}_{i}(||\widetilde{X}_{i}-\widetilde{X}_{i}(\tau_{p}^{i})||^{2}+\overline{\varepsilon}_{i0}^{2})\\
& -\int_{t-\mathcal{T}_{i}}^{t}||\dot{\widetilde{X}}_{i}(s)||^{2}ds,
\end{align*}
where $c_{4}$ is a positive constant.

For $\mathcal{V}_{i5}$, by (37) we have
\begin{align*}
\dot{\mathcal{V}}_{i5}\leq & c_{5}\mathcal{T}_{i}(||\widetilde{X}_{i}||^{2}+||\hat{X}_{i}||^{2}+\hat{w}^{i}(0,t)^{2})\\
& +c_{5}\mathcal{T}_{i}(||\widetilde{X}_{i}-\widetilde{X}_{i}(\tau_{p}^{i})||^{2}+||\hat{X}_{i}-\hat{X}_{i}(\tau_{p}^{i})||^{2})\\
& +c_{5}\mathcal{T}_{i}\overline{\varepsilon}_{i0}^{2}-\int_{t-\mathcal{T}_{i}}^{t}||\dot{\hat{X}}_{i}(s)||^{2}ds,
\end{align*}
where $c_{5}$ is a positive constant.

Summarizing the above inequalities, the derivative of $V_{i}$ is
given by:
\begin{align*}
\dot{\mathcal{V}}_{i}\leq & \sum_{j=1}^{3}\overline{\Lambda}_{ij}+\overline{\varepsilon}_{i},
\end{align*}
where $\overline{\varepsilon}_{i}\in\mathrm{\mathbb{E}}(0)$,
\begin{align*}
\overline{\Lambda}_{i1}= & \left(L_{i}+\ell+1+\lambda_{2}c_{2}+\lambda_{3}c_{3}\right.\\
& \left.+\mathcal{T}_{i}\lambda_{4}c_{4}+\mathcal{T}_{i}\lambda_{5}c_{5}\right)||\widetilde{X}_{i}||^{2}\\
& +(K_{i}+\ell+1+\mathcal{T}_{i}\lambda_{5}c_{5})||\hat{X}_{i}||^{2},
\end{align*}
\begin{align*}
\overline{\Lambda}_{i2}= & -\left(\lambda_{3}-c_{2}\lambda_{2}-c_{5}\mathcal{T}_{i}\lambda_{5}\right)\hat{w}^{i}(0,t)^{2},
\end{align*}
\begin{align*}
\overline{\Lambda}_{i3}=- & (\lambda_{4}-c_{1}\mathcal{T}_{i}\lambda_{1}-c_{2}\mathcal{T}_{i}\lambda_{2}-c_{3}\mathcal{T}_{i}\lambda_{3}\\
& -c_{4}\mathcal{T}_{i}\lambda_{4}-c_{5}\mathcal{T}_{i}\lambda_{5})\int_{t-\mathcal{T}_{i}}^{t}||\dot{\widetilde{X}}_{i}(s)||^{2}ds\\
- & (\lambda_{5}-c_{2}\mathcal{T}_{i}\lambda_{2}-c_{3}\mathcal{T}_{i}\lambda_{3}\\
& -c_{5}\mathcal{T}_{i}\lambda_{5})\int_{t-\mathcal{T}_{i}}^{t}||\dot{\hat{X}}_{i}(s)||^{2}ds.
\end{align*}
We can see that there exist small enough $\mathcal{T}_{i}$ and $\lambda_{j}(j=2,...,5)$
such that $\dot{\mathcal{V}}_{i}\leq-\overline{a}\mathcal{V}_{i}+\overline{\varepsilon}_{i}$.
The proof is completed.

\subsection{\textcolor{black}{Extension to high-order systems}}

We have mainly considered the control of a first-order nonlinear system
given by (3)-(5). However, the proposed control
scheme can be easily extended to more general nonlinear system, such
as strict-feedback nonlinear systems. We take the following high-order
nonlinear system as an example to show this.
\begin{align}
\dot{X}_{i1} & =X_{i2},\nonumber \\
\dot{X}_{i2} & =X_{i3}\nonumber \\
\cdots & \cdots\nonumber \\
\dot{X}_{in} & =f_{i}(X_{i1},X_{i2},...,X_{in})+U_{i}(t-D_{i})\label{eq:67}\\
Y_{i} & =X_{i1},\nonumber \\
e_{i} & =Y_{i}-y_{0},\nonumber 
\end{align}
where $X_{i}=(X_{i1},X_{i2},...,X_{in})^{\mathrm{T}}\in\mathbb{R}^{n}$,
$U_{i}\in\mathbb{R}$, $Y_{i}\in\mathbb{R}$, $e_{i}\in\mathbb{R}$
represent the system states, control input, system output and regulation
error respectively. $D_{i}>0$ denotes the long input delay. It is
assumed that $f_{i}(X_{i1},X_{i2},...,X_{in})$ satisfies the global
Lipschitz condition.

Let
\begin{align}
\overline{X}_{i1} & =e_{i}=X_{i1}-Fv,\label{eq:37-1-1}\\
\overline{X}_{i2} & =\dot{\overline{X}}_{i1}=X_{i2}-FSv,\\
\cdots & \cdots\\
\overline{X}_{in} & =\dot{\overline{X}}_{i,n-1}=X_{in}-FS^{n-1}v,
\end{align}
Then, the system (\ref{eq:67}) can be transformed into
\begin{align}
\dot{\overline{X}}_{i1} & =\overline{X}_{i2},\nonumber \\
\dot{\overline{X}}_{i2} & =\overline{X}_{i3}\nonumber \\
\cdots & \cdots\nonumber \\
\dot{\overline{X}}_{in} & =f_{i}(X_{i1},X_{i2},...,X_{in})-FS^{n-1}v+U_{i}(t-D_{i})\nonumber \\
& =\overline{f}_{i}(\overline{X}_{i1},...,\overline{X}_{in},v)+\overline{u}^{i}(0,t)\label{eq:72}\\
e_{i} & =\overline{X}_{i1},\nonumber 
\end{align}
where $\overline{X}_{i}=(\overline{X}_{i1},\overline{X}_{i2},...,\overline{X}_{in})^{\mathrm{T}}\in\mathbb{R}^{n}$,
$\overline{f}_{i}(\overline{X}_{i1},...,\overline{X}_{in},v)\triangleq f_{i}(\overline{X}_{i1}+Fv,...,\overline{X}_{in}+FS^{n-1}v)-f_{i}(Fv,...,FS^{n-1}v)$,
$R_{i}(v,S)\triangleq FS^{n-1}v-f_{i}(Fv,...,FS^{n-1}v)$. 
\begin{align*}
u^{i}(x,t) & \triangleq U_{i}(t+(x-1)D_{i}),\\
\overline{u}^{i}(x,t) & \triangleq u^{i}(x,t)-R_{i}(\mathrm{e}^{SD_{i}x}v,S),x\in[0,1]
\end{align*}
represent the distributed inputs. It is also noted that due to $f_{i}(X_{i1},X_{i2},...,X_{in})$
is globally Lipschitz, there exists a constant $\ell$ such that $\overline{f}_{i}(\overline{X}_{i1},...,\overline{X}_{in},v)\leq\ell(|\overline{X}_{i1}|+|\overline{X}_{i2}|+\cdots+|\overline{X}_{in}|)$.
Meanwhile, (\ref{eq:72}) can be written in a compact form
\begin{equation}
\overline{X}_{i}=\overline{f}_{i}^{c}(\overline{X}_{i1},...,\overline{X}_{in},v)+U_{i}(t-D_{i})\label{eq:73}
\end{equation}
where $\overline{f}_{i}^{c}(\overline{X}_{i1},...,\overline{X}_{in},v)$
is a nonlinear function.

According to the above dynamics, we can predict the states. Let
\begin{align*}
\chi^{i}(x,t) & \triangleq\overline{X}_{i}(t+D_{i}x)\\
& =\overline{X}_{i}(t)+D_{i}\int_{0}^{x}\left(\overline{f}_{i}^{c}(\chi^{i}(y,t),\mathrm{e}^{SD_{i}y}v)+\overline{u}^{i}(y,t)\right)dy.
\end{align*}
Then, the input $\overline{u}^{i}(1,t)$ is designed as:
\begin{align*}
& \overline{u}^{i}(1,t)\\
= & U_{i}-R_{i}(\mathrm{e}^{SD_{i}}v,S)\\
\triangleq & K_{i}\overline{X}_{i}(t+D_{i})\\
= & K_{i}\chi^{i}(0,t)+K_{i}D_{i}\int_{0}^{1}\left(\overline{f}_{i}^{c}(\chi^{i}(y,t),\mathrm{e}^{SD_{i}y}v)+\overline{u}_{i}(y,t)\right)dy,
\end{align*}
where $K_{i}=(k_{i1},k_{i2},...,k_{in}\mathrm{)}$ is the control
gain. The controller $u^{i}(1,t)=U_{i}(t)$ is then computed as
\begin{align}
& u^{i}(1,t)=U_{i}(t)\nonumber \\
= & K_{i}\overline{X}_{i}(t+D_{i})+R_{i}(\mathrm{e}^{SD_{i}}v,S)\nonumber \\
= & K_{i}D_{i}\int_{0}^{1}\left(\overline{f}_{i}(\chi_{i}(y,t),\mathrm{e}^{SD_{i}y}v)+\overline{u}_{i}(y,t)\right)dy\nonumber \\
& +K_{i}\overline{X}_{i}(t)+R_{i}(\mathrm{e}^{SD_{i}}v,S).\label{eq:35-3}
\end{align}
It follows that
\begin{align*}
\overline{u}^{i}(0,t) & =u^{i}(0,t)-R_{i}(v,S)\\
& =U_{i}(t-D_{i})-R_{i}(v,S)\\
& =K_{i}\overline{X}_{i}(t)+R_{i}(\mathrm{e}^{SD_{i}}v(t-D_{i}),S)-R_{i}(v,S)\\
& =K_{i}\overline{X}_{i}(t).
\end{align*}
This implies that (\ref{eq:73}) becomes
\begin{align*}
\dot{\overline{X}}_{i} & =\overline{f}_{i}^{c}(\overline{X}_{i},v)+K_{i}\overline{X}_{i}.
\end{align*}
It is noted that since $\overline{f}_{i}^{c}(\overline{X}_{i},v)$
satisfies the linear growth condition, we can always find a control
gain $K_{i}=(k_{i1},k_{i2},...,k_{in}\mathrm{)}$ such that the above
system is asymptomatic stable. That is $e_{i}(t)$
will converge to zero. 

Next, we will re-design the controller by the certainty equivalence
principle. The output of the controller $U_{i}(t)$ is designed as
\begin{align}
& U_{i}(t)\nonumber \\
= & K_{i}\hat{\chi}^{i}(0,t)\nonumber \\
& +K_{i}\hat{D}_{i}\int_{0}^{1}\left(\overline{f}_{i}(\hat{\chi}^{i}(y,t),\mathrm{e}^{\hat{S}_{i}\hat{D}_{i}y}\hat{S}_{i}\hat{v}_{i})+\hat{u}^{i}(y,t)\right)dy\nonumber \\
& +R_{i}(\mathrm{e}^{\hat{S}_{i}\hat{D}_{i}}\hat{v}_{i},\hat{S}_{i}),\label{eq:45-1-1}
\end{align}
where the notations are similar to (20). 

According to the above analysis, we have
\begin{thm}
	\label{thm:1-1-1-1-2} Given the MASs with leader (1)-(2)
	and nonlinear followers (\ref{eq:67}). There exists a control law
	(\ref{eq:45-1-1}) with periodic event-triggered adaptive distributed
	observer (6)-(7) such that Problem 1
	is solvable. Specifically, there exists a sufficiently small positive
	constant $D^{*}$ such that for $\forall|\hat{D}_{i}-D_{i}|\leq D^{*}(i=1,2,..,N)$,
	the output regulation error $|e_{i}(t)|\in\mathrm{\mathbb{E}}(\widetilde{D}_{i})$.
\end{thm}
\begin{proof}
	The proof follows the line of the main results in Section IV-B.
\end{proof}
\begin{rem}
	By following the same procedures, our proposed control scheme can
	be extended to strict-feedback nonlinear systems. Consider the following
	strict-feedback system
	\begin{align}
	\dot{X}_{i1} & =X_{i2}+f_{i1}(X_{i1}),\nonumber \\
	\dot{X}_{i2} & =X_{i3}+f_{i2}(X_{i1},X_{i2}),\nonumber \\
	\cdots & \cdots\nonumber \\
	\dot{X}_{in} & =f_{in}(X_{i1},X_{i2},...,X_{in})+U_{i}(t-D_{i}),\label{eq:67-1}\\
	Y_{i} & =X_{i1},\nonumber \\
	e_{i} & =Y_{i}-y_{0},\nonumber 
	\end{align}
	where $f_{ij}(X_{i1},X_{i2},..,X_{ij})$ are nonlinear function satisfying
	the global Lipschitz condition. Moreover, if we consider semi-global
	stability of the closed-loop system, the global Lipschitz condition
	can also be removed.
\end{rem}

\end{document}